\documentclass[11pt,a4paper]{article}

\usepackage{amsmath}
\usepackage{mathtools}
\usepackage{nccmath}
\usepackage{amsthm}
\usepackage{amssymb}
\usepackage{amsfonts}
\usepackage{anyfontsize}
\usepackage{mathrsfs}
\usepackage{graphicx}
\usepackage{subcaption}
\usepackage{caption}
\usepackage{cite}
\usepackage{bookmark,hyperref}
\usepackage{12many}
\usepackage{parskip}
\usepackage{setspace}
\usepackage{multirow}
\usepackage{arydshln}
\usepackage{anyfontsize}
\usepackage{changepage}
\usepackage{todonotes}
\usepackage{braket}
\usepackage{dirtytalk}
\usepackage{MnSymbol}%This one changes the font of arrows, the infty symbol and so on.
\usepackage{tikz-cd}
\usepackage{adjustbox}
\usepackage{bm}

\newcommand{\widebar}[1]{\mskip.5\thinmuskip\overline{\mskip-.5\thinmuskip {#1} \mskip-.5\thinmuskip}\mskip.5\thinmuskip} % overline short

\newcommand{\rmd}{\mathrm{d}}

\numberwithin{equation}{section}
\usepackage{fullpage}

\begin{document}
\begin{titlepage}
 \thispagestyle{empty}
 \begin{flushright}
 \hfill{Imperial-TP-2025-CH-4 }\\
   
 \end{flushright}

 \vspace{30pt}

 \begin{center}
     
  {\fontsize{20}{24} \bf {Duality and Infinite Distance Limits in\\[7mm] Asymmetric Freely Acting Orbifolds}}

     \vspace{30pt}
{\fontsize{13}{16}\selectfont {George Gkountoumis$^1$, Chris Hull$^2$, Guo-En Nian$^1$ and Stefan Vandoren$^1$}} \\[10mm]

{\small\it
${}^1$ Institute for Theoretical Physics {and} Center for Extreme Matter and Emergent Phenomena \\
Utrecht University, 3508 TD Utrecht, The Netherlands \\[3mm]

${}^2$ %Imperial College, London, UK \\[3mm]
{{The Blackett Laboratory}},
{{Imperial College London}}\\
{{Prince Consort Road}}, 
{{London SW7 2AZ, U.K.}}\\[3mm]}

\vspace{1.5cm}

{\bf Abstract}

\vspace{0.3cm}
   
\begin{adjustwidth}{12pt}{12pt}

We use freely acting asymmetric orbifolds of type IIB string theory to construct a class of theories in four dimensions with eight supercharges. Their low energy effective field theories resemble $STU$ models, but have different duality groups: the orbifold's free action reduce the duality groups to congruence subgroups of the modular group. The fundamental domain is consequently  larger and contains new interesting points at infinite distance on the real axis bounding the upper half plane. We
verify that the   distance conjectures hold in the non-geometric compactification of string theory studied here. In particular, we find points at infinite distance in moduli space at which the theory decompactifies to a different orbifold construction.

\end{adjustwidth}

\end{center}

\vspace{20pt}

\newcommand\blfootnote[1]{%
  \begingroup
  \renewcommand\thefootnote{}\footnote{#1}%
  \addtocounter{footnote}{-1}%
  \endgroup}

\blfootnote{g.gkountoumis@uu.nl \quad c.hull@imperial.ac.uk \quad g.nian@uu.nl \quad s.j.g.vandoren@uu.nl}

\noindent

\end{titlepage}

\begin{spacing}{1.15}
\tableofcontents
\end{spacing}

\section{Introduction}

Recently, there has been a revival of interest in  asymmetric orbifolds \cite{bianchi2022perturbative,Gkountoumis:2023fym, Baykara:2023plc,Gkountoumis:2024dwc, Gkountoumis:2024boj,Baykara:2024vss,Baykara:2024tjr,Angelantonj:2024jtu,Aldazabal:2025zht,Baykara:2025lhl}. In general, toroidal orbifolds provide a way to construct string models with reduced (or no) supersymmetry, and asymmetric orbifolds are particularly interesting because many more moduli can be frozen for these models than can be frozen for   symmetric orbifolds. If we orbifold by a  freely acting symmetry, then all states in the twisted sectors are typically massive at  generic points in the moduli space, and hence  the number of moduli is reduced even further. If the free  action involves a shift on a circle coordinate, the orbifold can also be understood as a string compactification with a duality twist \cite{Dabholkar:2002sy}. Asymmetric orbifolds constitute an interesting region in the string landscape of non-geometric string compactifications, and their corresponding effective supergravity theories provide highly non-trivial examples, in which the various conjectures of the swampland programme\cite{vafa2005string} (see \cite{Palti:2019pca,Agmon:2022thq} for a review and lecture notes) can be tested.

As we break supersymmetry, using toroidal orbifolds, the duality group  $G(\mathbb{Z})$ arising from toroidal compactification is broken by the orbifold twist to a subgroup $K(\mathbb{Z})\subset G(\mathbb{Z})$, which is a symmetry of the untwisted sector. When the orbifold twist is accompanied by a shift on a circle, the duality group of the untwisted sector reduces to an even smaller subgroup $\Gamma(\mathbb{Z})\subset K(\mathbb{Z})$, which is the group that  preserves both the orbifold twist and the shift. This breaking of the toroidal duality group was seen in \cite{sen1995dual} and  we revisit and  further discuss it in this paper. There are interesting and  important consequences for the moduli space of such orbifolds, since we need to quotient by $\Gamma(\mathbb{Z})$ to get the space of inequivalent theories. The moduli space then takes the form $\tilde{\cal M}={\cal M}/\Gamma(\mathbb{Z})$ and should obey all the swampland conjectures such as the distance conjecture and the finiteness of the  volume of $\tilde{\cal M}$ \cite{Ooguri:2006in}, see also \cite{Grimm:2018ohb,Corvilain:2018lgw} for more on the distance conjecture in the context of Calabi-Yau compactifications.

Freely acting orbifolds resemble models with spontaneous supersymmetry breaking (see e.g. \cite{Rohm:1983aq,Ferrara:1987es,Kounnas:1988ye,
Ferrara:1988jx}), as some (or all) gravitini obtain masses instead of being projected out of the orbifold spectrum. In this work, we focus on orbifolds breaking supersymmetry spontaneously from $\mathcal{N}=8$ (32 supersymmetries) to $\mathcal{N}=2$ (8 supersymmetries) in $4D$. At points of infinite distance in the moduli space, some (or all) gravitini can become massless, indicating supersymmetry enhancement. For the models studied in this work, all such points can be understood as decompactification limits of the original orbifold theory.
Remarkably, at points on the real axis of the upper half plane, an asymmetric freely acting orbifold  may decompactify to a non-freely acting symmetric orbifold. We will discuss such cases in detail here.

In section \ref{FAO} we review the general construction of asymmetric freely acting orbifolds of type IIB on $T^6$, and we discuss the S and T-duality groups of the  orbifold theory, as well as those of the effective supergravity theory. Then in section \ref{z12 with nv=3 and nh=0} we construct an $STU$-like model by using an asymmetric freely acting $\mathbb{Z}_6$ orbifold, preserving 8 supersymmetries in four dimensions, and we analyse the spectrum of lightest states in the untwisted and twisted orbifold sectors. In section \ref{moduli and swampland} we discuss the classical moduli space of our orbifold model and we carefully determine all special points and lines in the moduli space where generically massive states can become massless. At infinite distance points in the moduli space we find infinite towers of states becoming massless. In the bulk of the moduli space we find only a finite number of states that become massless. We conclude with a discussion in section \ref{Conclusion and discussion}.

\section{Freely acting orbifolds}\label{FAO}
In this section we first briefly review freely acting asymmetric orbifolds of type IIB string theory on $T^6$, closely following a similar discussion in \cite{Gkountoumis:2023fym,Gkountoumis:2024dwc}. We then discuss the S and T-duality groups of the orbifold theory, and those of the effective supergravity theory.

\subsection{General remarks}
\label{general remarks}
The orbifolds we focus on here  have target spaces of the form 
\begin{equation}\label{background}
\mathbb{R}^{1,3}\times  T^4\times T^2\ , \end{equation}
identified under the action of a $\mathbb{Z}_p$ symmetry.
At a point in moduli space where the  $T^4$ CFT has a $\mathbb{Z}_p$ symmetry, we orbifold by this $\mathbb{Z}_p$ acting on the $T^4$ CFT combined with a shift along one cycle of the $T^2$ which makes the orbifold freely acting.
Freely acting orbifolds have no fixed points and, at generic points in the moduli space, all states coming from the twisted sectors are massive. 
Supersymmetry is   spontaneously broken in these models with some (or all) of the gravitini becoming massive, in contrast to non-freely acting orbifolds
in which some (or all) gravitini are projected out and supersymmetry is  explicitly broken.

Symmetric orbifolds arise when the $\mathbb{Z}_p$ action on $T^4$ is a geometric discrete symmetry of $T^4$, generated by a diffeomorphism on $T^4$, i.e. by an element in GL(4;$\mathbb{Z}$). These orbifolds preserve $\mathcal{N}=4$ or $0$ supersymmetry in $4D$. For asymmetric orbifolds, the $\mathbb{Z}_p$ group acts as a T-duality transformation on the $T^4$ CFT. The T-duality group for superstrings on $T^4$ is Spin$(4,4;\mathbb{Z})$, a discrete subgroup
of the double cover Spin$(4,4)$ of $\text{SO}(4,4)$, as the D-brane states transform as a spinor representation of $\text{Spin}(4,4)$ \cite{Hull:1994ys}. The background fields, namely the  torus metric $G$ and the two-form $B$-field, can be combined  into a matrix $E=G+B$. T-duality transforms $E$ to a new background $E'$ through a fractional linear transformation\footnote{For details on how T-duality acts on the background fields we refer to the classic review \cite{giveon1994target} and e.g. \cite{tan2015t,satoh2017lie}. }. Consistency of the asymmetric orbifold then requires that the $\mathbb{Z}_p$ transformation is a symmetry under which $E'=E$. This can be achieved only for special values of the  moduli, which are therefore stabilised in these non-geometric constructions. The class of asymmetric orbifolds we consider preserve $\mathcal{N}=6,4,2$ or $0$ supersymmetry in $4D$.

\subsection{Asymmetric orbifolds}
\label{asymmetric orbifold}

For the construction of asymmetric orbifolds we follow the procedure presented in the original papers \cite{Narain:1986qm,narain1991asymmetric}. In general, upon compactification of the IIB superstring on $T^6$ the momentum and winding numbers take values in a Narain lattice $\Gamma^{6,6}$, which is an even, self-dual lattice \cite{narain1989new}. For our purposes we decompose $T^6=T^4\times T^2$ and correspondingly decompose the lattice as $\Gamma^{4,4}(\mathcal{G})\oplus \Gamma^{2,2}$, where $\Gamma^{4,4}(\mathcal{G})$ is an even, self-dual Lie algebra lattice 
which has symmetries acting purely on the left-movers and symmetries acting purely on the right-movers.
Such a lattice can be realised at special points in the moduli space as 
\begin{equation}
    \Gamma^{4,4}(\mathcal{G}) \equiv \{(p_{\text{L}},p_{\text{R}})|\,p_{\text{L}} \in \Lambda_W(\mathcal{G}),\, p_{\text{R}} \in \Lambda_W(\mathcal{G}),\, p_{\text{L}}-p_{\text{R}} \in \Lambda_{\text{R}}(\mathcal{G})\}\,.
\end{equation}
Here $\mathcal{G}$ is a Lie algebra of rank four and $\Lambda_W(\mathcal{G})$, $\Lambda_{\text{R}}(\mathcal{G})$ denote the weight and root lattices of $\mathcal{G}$, respectively. Now, the symmetry we orbifold by acts as an automorphism of   $\Gamma^{4,4}(\mathcal{G})$ and as a shift on $\Gamma^{2,2}$. Here we will only consider automorphisms  that act as a rotation on the left-movers and a separate rotation on the right-movers\footnote{A discussion on consistent rotations can be found e.g.\ in \cite{lerchie1989lattices}, cf. appendix B.}, $\mathcal{M}_{\theta}=(\mathcal{N}_{\text{L}},\mathcal{N}_{\text{R}}) \in \text{SO}(4)_{\text{L}}\times \text{SO}(4)_{\text{R}} \subset \text{SO}(4,4)\,$, the latter being the T-duality group of $T^4$. 
For a $\mathbb{Z}_p$ orbifold, we require that the rotation satisfies 
$(\mathcal{M}_{\theta})^p=1$. Also, in order to properly define the orbifold action on fermions the rotation matrix $\mathcal{M}_{\theta}\in \text{SO}(4,4)$ should be uplifted to a {monodromy} matrix $\mathcal{M}\in \text{Spin}(4,4)$; for more details see \cite{Gkountoumis:2023fym}.

Since the orbifold acts as a shift on $\Gamma^{2,2}$, it leaves $\Gamma^{2,2}$ invariant\footnote{Due to the shift, momentum states pick up a phase in the untwisted sector. In the twisted sectors states become massive.}. On the other hand, $\Gamma^{4,4}(\mathcal{G})$ is not in general invariant under the rotation $\mathcal{M}_{\theta}$. The sublattice $I\subset \Gamma^{4,4}(\mathcal{G})$ that is invariant under the orbifold action is given by
\begin{equation}
    I \equiv \{ p \in \Gamma^{4,4}(\mathcal{G})\,|\, \mathcal{M}_{\theta}\cdot p = p\}\,.
\end{equation}
Then the complete  sublattice that is invariant under the 
orbifold action
is 
\begin{equation}
    \hat{I}=I\oplus \Gamma^{2,2}\,.
\end{equation}
It will be useful to determine the orbifold action on the world-sheet fields. We denote the two real $T^2$
coordinates by $Z_1, Z_2$ and the  four real $T^4$ coordinates by $Y^m, m=1,\ldots 4$, which we combine into two complex coordinates
 $W^i = \tfrac{1}{\sqrt{2}}(Y^{2i-1}+iY^{2i})$ with $i=1,2$. We parametrize the rotations $(\mathcal{N}_{\text{L}},\mathcal{N}_{\text{R}}) \in \text{SO}(4)_{\text{L}}\times \text{SO}(4)_{\text{R}} \subset \text{SO}(4,4)\,$ by four mass parameters 
\begin{equation}
    \mathcal{N}_{\text{L}} = \begin{pmatrix}
        R(m_1+m_3)&0\\
        0&R(m_1-m_3)
        \end{pmatrix}\,, \qquad \mathcal{N}_{\text{R}} = \begin{pmatrix}
        R(m_2+m_4)&0\\
        0&R(m_2-m_4)
        \end{pmatrix}\,,
        \label{rotations in terms of mass parameters}
\end{equation}
where we use the notation $R(x)=\begin{psmallmatrix}\cos x & \,\,\,\,-\sin x \\ \sin x & \,\,\,\,\cos x \end{psmallmatrix}$ for a $2\times 2$ rotation matrix. Then, the orbifold acts on the bosonic torus coordinates through (asymmetric) rotations
\begin{equation}\label{orbiaction2}
\begin{aligned}
{W}_{\text{L}}^1 \;&\rightarrow\; e^{i(m_1+m_3)}\: {W}_{\text{L}}^1 \,, \\
{W}_{\text{L}}^2 \;&\rightarrow\; e^{i(m_1-m_3)}\: {W}_{\text{L}}^2 \,, \\
W_{\text{R}}^1 \;&\rightarrow\; e^{i(m_2+m_4)}\: W_{\text{R}}^1 \,, \\
W_{\text{R}}^2 \;&\rightarrow\; e^{i(m_2-m_4)}\: W_{\text{R}}^2 \,,
\end{aligned}
\end{equation}
and with the same action on their world-sheet superpartners. Symmetric orbifolds arise in the case in which $m_1=m_2$ and $m_3=m_4$. The rotations on the torus are accompanied by a shift along one of the $T^2$ coordinates, which makes the orbifold freely acting. Without loss of generality, we choose
\begin{equation}\label{shift}
\begin{aligned}
&Z_1 \;\rightarrow\; Z_1 + 2\pi \mathcal{R}_5 / p \,,\\
&Z_2 \;\rightarrow\; Z_2\,,
\end{aligned}
\end{equation}
where $\mathcal{R}_5$ is the radius of the $S^1$ on which the orbifold acts as a shift. Also, we will denote by $\mathcal{R}_4$ the radius of the $S^1$ which is inert under the orbifold action. 

In order for strings to close in our geometry, they need to satisfy the boundary conditions
\begin{equation}\label{boundaryconditions}
\begin{alignedat}{4}
X^{\hat\mu}(\sigma^1, \sigma^2+2\pi) &= X^{\hat{\mu}}(\sigma^1, \sigma^2) \,, \quad
&Z_1(\sigma^1, \sigma^2+2\pi) &= Z_1(\sigma^1, \sigma^2) + 2\pi \mathcal{R}_5(w^1 + {k}/{p}) \,, \\[4pt]
W_{\text{L}}^1(\sigma^1, \sigma^2+2\pi) &= e^{ik(m_1+m_3)} \:W_{\text{L}}^1(\sigma^1, \sigma^2) \,, \quad &W_{\text{R}}^1(\sigma^1, \sigma^2+2\pi) &= e^{ik(m_2+m_4)} \:W_{\text{R}}^1(\sigma^1, \sigma^2) \,, \\[4pt]
W_{\text{L}}^2(\sigma^1, \sigma^2+2\pi) &= e^{ik(m_1-m_3)} \:W_{\text{L}}^2(\sigma^1, \sigma^2) \,, \quad &W_{\text{R}}^2(\sigma^1, \sigma^2+2\pi) &= e^{ik(m_2-m_4)} \:W_{\text{R}}^2(\sigma^1, \sigma^2) \,.
\end{alignedat}
\end{equation}
Here, $X^{\hat{\mu}}=\{X^0,\ldots, X^3, Z_2\}$, $\sigma^1$ and $\sigma^2$ are the coordinates on the worldsheet, which we always take to be of Lorentzian signature, and $w^1 \in \mathbb{Z}$ is the winding number along the $S^1$ on which the orbifold acts with a shift (we omit writing down the winding modes on the rest of the compact directions here for simplicity of the formulae). Also, $k = 0,\ldots,p-1$ is an integer that distinguishes between the various sectors. We have the untwisted sector for $k=0$, and $p-1$ twisted sectors for the other values of $k$ in which case the string closes up to the action of the $\mathbb{Z}_p$ symmetry.

For each mass parameter that is not zero (mod $2\pi$) a pair of gravitini becomes massive. So, in order to obtain an $\mathcal{N}=2$ theory only one mass parameter should be set to zero. These mass parameters can be translated to the more familiar language of twist vectors used in the orbifold literature as follows\footnote{Comparing with the notation used in \cite{Gkountoumis:2023fym}, we simply omit here the two first trivial entries of the twist vectors. Comparing with \cite{Baykara:2023plc}, we have $\tilde{u}=\phi_{\text{L}}$ and $u=\phi_{\text{R}}$.}
\begin{equation}
\begin{aligned}
   & \tilde{u} \equiv (\tilde{u}_3,\tilde{u}_4)= \frac{1}{2\pi}\left(m_1+m_3,m_1-m_3\right)\ ,\\
   &u\equiv(u_3,u_4)=\frac{1}{2\pi}\left(m_2+m_4,m_2-m_4\right)\ .
    \end{aligned}
    \label{relation twist vectors-mass parameters}
\end{equation}
With these at hand, it is straightforward to see that
\begin{equation}
    \mathcal{N}_{\text{L}} = \begin{pmatrix}
        R(2\pi\tilde{u}_3)&0\\
        0&R(2\pi\tilde{u}_4)
        \end{pmatrix}\,, \qquad \mathcal{N}_{\text{R}} = \begin{pmatrix}
        R(2\pi u_3)&0\\
        0&R(2\pi u_4)
        \end{pmatrix}\,,
        \label{orbifold twist in terms of twist vectors}
\end{equation}
Also, the shift along the one circle of $T^2$ \eqref{shift} can be represented by a shift vector 
\begin{equation}
    v=  \begin{pmatrix}
  \frac{1}{p}\\
   0 \\
   0\\
   0
    \end{pmatrix}\,.
    \label{shiftvapp}
\end{equation}
This is in the vector representation of ${\rm Spin}(2,2)$ and written in the basis of winding $(w^i)$ and momentum $(n_i)$ numbers. Recall that in this basis a vector of the lattice $\Gamma^{2,2}$, associated with $T^2$, can be written as
\begin{equation}\label{latticeP}
    P=  \begin{pmatrix}
  w^1\\
   w^2 \\
   n_1\\
   n_2
    \end{pmatrix}\,.
\end{equation}

Finally, we have to ensure that our models are modular invariant. This requires that the following conditions hold \cite{vafa1986modular,Baykara:2023plc}
\begin{equation}
    p \sum_{i=3}^4 \tilde{u}_i \,\in\, 2\mathbb{Z} \qquad\text{and}\qquad p \sum_{i=3}^4 u_i \,\in\, 2\mathbb{Z} \ ,
    \label{modular conditions on twist vectors}
\end{equation}
where $p$ is the orbifold rank. Also, if $p$ is even, we need to check the additional condition for the momenta\footnote{In some cases it is possible to construct consistent orbifolds even if this condition is not satisfied \cite{Harvey:2017rko}.} $(p_{\text{L}},p_{\text{R}})\in \Gamma^{4,4}(\mathcal{G})$
\begin{equation}\label{modcond2}
    p_{\text{L}} \mathcal{N}_{\text{L}}^{\,p/2} p_{\text{L}} - p_{\text{R}} \mathcal{N}_{\text{R}}^{\,p/2} p_{\text{R}} \,\in\, 2\mathbb{Z}\,. 
\end{equation}

\subsection{Duality groups of the orbifolded theory}
\label{duality groups of orbifolded theory}

The compactification of type II string theory on $T^6$ has $\mathcal{N}=8$ supergravity as its low energy effective field theory. This has  70 massless scalars that parametrise the moduli space
\begin{equation}
    {\cal M}=\frac{\text{E}_{7(7)}}{\text{SU}(8)/\mathbb{Z}_2}\ .
\end{equation}
The 38 scalars from the NS-NS sector parametrise a subspace, which factorises as
\begin{equation}
    {\cal M}_{\text{NS}}=\frac{\text{SL}(2)}{\text{U}(1)}\times \frac{\text{Spin}(6,6)}{\text{Spin}(6)\times \text{Spin}(6)}\ .
\end{equation}
In type IIB, the first factor is parametrised by the axion and a shifted dilaton (the axion is a scalar dual to the $4D$ NS-NS two form) and the second factor is parameterised by the 36 moduli for the metric and $B$-field  on $T^6$.

The duality group of the effective supergravity theory is $\text{E}_{7(7)}$, which has a maximal subgroup
\begin{equation}
    \text{E}_{7(7)}\supset \text{SL}(2)\times \text{Spin}(6,6)\ .
\end{equation}
In the quantum theory the  duality group $\text{E}_{7(7)}$ is broken to its discrete U-duality subgroup $\text{E}_{7(7)}(\mathbb{Z})$, so that the 
$\text{SL}(2)\times \text{Spin}(6,6)$  subgroup is broken to a discrete group 
$\text{SL}(2;\mathbb{Z})\times \text{Spin}(6,6;\mathbb{Z})$, forming the S and T-duality groups of type II string theory on $T^6$ \cite{Hull:1994ys}. We mention that this $4D$ S-duality, which will be referred to here as $\text{SL}(2)_S$,  should not be confused with the $10D$ S-duality of type IIB.
Note that the {\bf 56} representation of $\text{E}_{7(7)}$ decomposes into 
$\text{SL}(2)\times \text{Spin}(6,6)$ representations as follows:
\begin{equation}
\textbf{56}\to (\textbf{2},\textbf{12})+(\textbf{1},\textbf{32}).
\label{ 56dec}
\end{equation}
The NS-NS charges are in the $(\textbf{2},\textbf{12})$ representation, consisting of the 6 momenta and 6 winding numbers, which are the perturbative charges, together with 6 NS5-brane charges and 6 KK monopole charges, which are the non-perturbative charges \cite{Hull:1994ys}.

Orbifolding this theory  breaks the U-duality group $\text{E}_{7(7)}(\mathbb{Z})$ to a subgroup. The orbifold is specified by a twist and a shift, and the  U-duality group is broken to the subgroup that preserves both of these, which we will refer to as the orbifold duality group. This is then the subgroup of $\text{E}_{7(7)}(\mathbb{Z})$ that commutes with the twist and preserves the shift vector (up to the addition of a lattice vector and up to a sign; see below).

For the theories we consider in this paper with decomposition $T^6=T^4\times T^2$, the orbifold duality group is in fact a subgroup of $\text{SL}(2;\mathbb{Z})\times \text{Spin}(6,6;\mathbb{Z})$.
The subgroup of $\text{Spin}(6,6) $ acting as T-duality on $T^4$ is $\text{Spin}(4,4) $ and, for each orbifold, the twist matrix specified by \eqref{orbifold twist in terms of twist vectors} is in the compact part of this, 
$\text{Spin}(4)\times \text{Spin}(4) $. 
Let $\mathcal{C}$ be the subgroup of
${\rm Spin}(4,4)$ that commutes with the twist. 
For the theories considered here, the twist then breaks
${\rm Spin}(6,6)$ to ${\rm Spin}(2,2)\times \mathcal{C}$ in the supergravity theory. In the orbifolded string theory, the group $\mathcal{C}$ is broken further to a discrete subgroup $\mathcal{C}(\mathbb{Z})$.

The shift vector \eqref{shiftvapp} leads to a further breaking of the symmetry. As in \eqref{shift}, this vector $v$ represents an orbifold shift along the circle with radius $\mathcal{R}_5$ by $2\pi \mathcal{R}_5/p$. Adding any integer-valued 4-vector 
$w$ in the lattice $ \mathbb{Z}^4$
to $v$  will give the same orbifold.
Furthermore, $v$ and $-v$ specify physically equivalent orbifolds as the sign of $v$ is changed by the reflection $Z_1\to -Z_1$; such a reflection is an element of the T-duality group $\text{Spin}(6,6,\mathbb{Z})$. Note that changing $v$ to $-v$ has the same effect as replacing the monodromy
$\mathcal{M}$ with $\mathcal{M}^{-1}$. If $\mathcal{M}$ is expressed in terms of a mass matrix $M$ by $\mathcal{M}=e^{M}$, then this amounts to changing the sign of the mass matrix $M$.

Note that this discussion of duality symmetries applies to the untwisted sector, and we discuss the subgroup of the original duality that is a symmetry of the untwisted sector of the orbifold.  However, the orbifold introduces twisted sectors, and there can be new duality symmetries relating the untwisted and twisted sectors that do not directly arise from the duality symmetry of the theory before the orbifold.
For example, consider type IIA string theory compactified on $T^4$, with U-duality symmetry ${\rm Spin}(5,5;\mathbb{Z})$. Consider now the $\mathbb{Z}_2$ orbifold of this which gives a special point in the $\mathrm{K}3$ moduli space. The untwisted sector is invariant under an $\text{SO}(4,4;\mathbb{Z})$ subgroup of the duality group, but the full theory in fact has a duality symmetry $\text{SO}(4,20;\mathbb{Z})$. Note that $\text{SO}(4,20)$ is not a subgroup of ${\rm Spin}(5,5)$ and the extra duality symmetries include ones mixing untwisted and twisted sectors. Whether or not there are extra symmetries of this kind depends on the model.
We will see below that some of our orbifold examples have this kind of duality enhancement.

Before turning to the subgroup of ${\rm Spin}(2,2)$ that is preserved in the orbifold, 
we discuss an $\text{SL}(2)$ analogue that introduces the relevant groups.
Consider, then,  the simpler problem of finding
the subgroup of $\text{SL}(2;\mathbb{Z})$
that preserves a 2-vector of the form
\begin{equation}
    V=  \begin{pmatrix}
  \frac{1}{p}\\
   0
    \end{pmatrix}
    \label{shiftvvapp}
\end{equation}
up to the addition of a lattice vector $\begin{pmatrix}
  w\\
   n
    \end{pmatrix} \in \mathbb{Z}^2$. It is easy to check that the result is the group
\begin{equation}
  \Gamma_1(p) = \Bigg\{ \begin{pmatrix}
       a &b\\
        c& d
    \end{pmatrix}\in  \text{SL}(2;\mathbb{Z}):\quad a,d=1 \:\text{mod} \:\,p\,,\quad c=0\: \text{mod} \:\, p\Bigg\}\,.
    \label{gamma1}
\end{equation}
The subgroup preserving $V$ up to a sign, i.e.\ taking $V$ to $\pm V$ plus a lattice vector, is 
 \begin{equation}
  \hat \Gamma_1(p) = \Bigg\{ \begin{pmatrix}
       a &b\\
        c& d
    \end{pmatrix}\in  \text{SL}(2;\mathbb{Z}):\quad a,d=\pm 1 \:\text{mod} \:\,p\,,\quad c=0\: \text{mod} \:\, p\Bigg\}\,.
    \label{gamma1h}
\end{equation}

For $p=2$, we have that $-1=1$ mod $2$ and so 
\begin{equation}
    \hat \Gamma_1(2) =  \Gamma_1(2)\, ,
\end{equation}
 while for $p>2$ 
\begin{equation}
    \hat \Gamma_1(p) =  \Gamma_1(p) \times \mathbb{Z}_2\, ,
\end{equation}
 with the $\mathbb{Z}_2$ consisting of the $2\times 2 $ matrices 
 $\{ \bm{1},-\bm{1} \}$.
 
 For $\text{SL}(2;\mathbb{R})$, the subgroup preserving $V$ up to a lattice vector is 
 \begin{equation}
  \Delta_1(p) = \Bigg\{ \begin{pmatrix}
       a &b\\
        c& d
    \end{pmatrix}\in  \text{SL}(2;\mathbb{R}):\quad a=1 \:\text{mod} \:\,p\,,\quad c=0\: \text{mod} \:\, p\Bigg\}\, ,
    \label{Del1}
\end{equation}
while the subgroup taking $V$ to $\pm V$ plus a lattice vector is  
 \begin{equation}
 \hat  \Delta_1(p) = \Bigg\{ \begin{pmatrix}
       a &b\\
        c& d
    \end{pmatrix}\in  \text{SL}(2;\mathbb{R}):\quad a=\pm 1 \:\text{mod} \:\,p\,,\quad c=0\: \text{mod} \:\, p\Bigg\}\, ,
    \label{Delh1}
\end{equation}
Both of these contain the stability subgroup preserving  $V$, which is the group $\mathbb{R}$ of upper triangular matrices
\begin{equation}
  \begin{pmatrix}
       1 &b\\
        0& 1
    \end{pmatrix}\,.
    \label{uptri}
\end{equation}
We now return to the breaking of the T-duality group of $T^2$, which is 
\begin{equation}
    {\rm Spin}(2,2;\mathbb{Z})\cong\text{SL}(2;\mathbb{Z})_T\times \text{SL}(2;\mathbb{Z})_U\,,
\end{equation}
where $\text{SL}(2;\mathbb{Z})_T$ acts on the complexified K\" ahler modulus $T$ of the $T^2$, while $\text{SL}(2;\mathbb{Z})_U$ acts on the complex structure modulus $U$.\footnote{For a review of T-duality see \cite{giveon1994target}.} We denote the (dimensionful) metric and antisymmetric Kalb-Ramond $B$-field on $T^2$ by $g_{ij}$ and $b_{ij}$, respectively. Here $i,j=1,2$, and  $b_{ij}=b\,\varepsilon_{ij}$, where $b$ is a constant, and $\varepsilon_{12}=-\varepsilon_{21}=1$, $\varepsilon_{11}=\varepsilon_{22}=0$. Given these, we can define the moduli $T$ and $U$ as (see e.g. \cite{Blumenhagen:2013fgp})
\begin{equation}
\begin{aligned}
  &T = T_1+i T_2 = \frac{1}{\alpha'}\left(b + i \sqrt{\text{det}g}\right) \,,\\
& U = U_1+iU_2 = \frac{g_{12}}{g_{11}} + i \frac{\sqrt{\text{det}g}}{g_{11}} \,.
 \end{aligned}
 \label{T,Udefap}
\end{equation}
Then, the $T^2$ metric can expressed in terms of these moduli as
\begin{equation}
    g_{ij} = \alpha'\frac{T_2}{U_2}\begin{pmatrix}
        1&U_1\\
        U_1 & |U|^2
    \end{pmatrix}\,.
\end{equation}
Note that for a rectangular torus, that is $U_1=0$, and a trivial $B$-field, $T_2=\mathcal{R}_5\mathcal{R}_4/\alpha'$ and $U_2=\mathcal{R}_4/\mathcal{R}_5$, where $\mathcal{R}_5$ is the radius of the circle on which the orbifold acts by a shift and $\mathcal{R}_4$ is the radius of the circle that is invariant under the orbifold action.

The $T^2$ partition function, in the untwisted orbifold sector and without the insertion of the orbifold group element, reads
\begin{equation}
{Z}_{T^2}[0,0]=  \frac{1}{(\eta\widebar{\eta})^2}\sum_{\left\{n_i,w_i\right\} \in \mathbb{Z}^4}\, \widebar{q}^{\frac{1}{2}p_{\text{L}}^2}\,q^{\frac{1}{2}p_{\text{R}}^2}\,, \qquad i=1,2\,,
\end{equation}
where\footnote{Here and below we label both momentum and winding numbers by a subscript for simplicity of the formulae.} 
\begin{equation}
    \begin{aligned}
        &p^2_{\text{L}} = \frac{1}{2T_2U_2}|n_2-Un_1 +\widebar{T}w_1 +\widebar{T}Uw_2|^2\,,\\
         &p^2_{\text{R}} = \frac{1}{2T_2U_2}|n_2-Un_1 +{T}w_1 +{T}Uw_2|^2\,.
    \end{aligned}
    \label{momentauntwisted}
\end{equation}
Now, the two $\text{SL}(2;\mathbb{Z})$ subgroups of the T-duality group of $T^2$ act on the moduli $T$ and $U$ as follows:
\begin{equation}
    \text{SL}(2;\mathbb{Z})_T: \quad\frac{aT+b}{cT+d}  \,,\qquad ad-bc=1\ ,
\end{equation}
\begin{equation}
   \quad \:\, \text{SL}(2;\mathbb{Z})_U: \quad\frac{a'U+b'}{c'U+d'}    \,,\quad \:\:a'd'-b'c'=1\,.
\end{equation}
The T-duality group also acts on the momentum and winding numbers by $\text{Spin}(2,2;\mathbb{Z})$ transformations, leaving $p^2_{\text{L/R}}$ invariant. The action on the Narain lattice is given by (see e.g. \cite{Bailin:1993wv})
\begin{equation}
     \text{SL}(2;\mathbb{Z})_T: \quad   \begin{pmatrix}
  w_1\\
   w_2 \\
   n_1\\
   n_2
    \end{pmatrix}\to  \begin{pmatrix}
  d&0&0&-c\\
   0&d&c&0 \\
   0&b&a&0\\
   -b&0&0&a
    \end{pmatrix}  \begin{pmatrix}
  w_1\\
   w_2 \\
   n_1\\
   n_2
   \end{pmatrix}\,,
   \label{sl2t}
\end{equation}
\begin{equation}
 \quad   \: \text{SL}(2;\mathbb{Z})_U: \quad   \begin{pmatrix}
  w_1\\
   w_2 \\
   n_1\\
   n_2
    \end{pmatrix}\to  \begin{pmatrix}
  a'&-b'&0&0\\
   -c'&d'&0&0 \\
   0&0&d'&c'\\
   0&0&b'&a'
    \end{pmatrix}  \begin{pmatrix}
  w_1\\
   w_2 \\
   n_1\\
   n_2
   \end{pmatrix}\,,
   \label{sl2u}
\end{equation}
Notice that $\text{SL}(2;\mathbb{Z})_U$ does not mix winding and momenta, but $\text{SL}(2;\mathbb{Z})_T$ does. It is easy to check that the two $\text{SL}(2)$'s commute in the basis of winding and momentum numbers. Also, an element $(g_T,g_U)\in\text{SL}(2;\mathbb{Z})_T \times \text{SL}(2;\mathbb{Z})_U$ can be embedded in $\text{SO}^+(2,2;\mathbb{Z})$ as\footnote{
$\text{SO}^+(2,2;\mathbb{Z})$ is the component of $\text{O}(2,2;\mathbb{Z})$ connected to the identity. Also, $\text{SL}(2;\mathbb{Z}) \times \text{SL}(2;\mathbb{Z}) \cong \text{Spin}(2,2;\mathbb{Z})$ is the double cover of $\text{SO}^+(2,2;\mathbb{Z})$.}
\begin{equation}
(g_T,g_U) \in\text{SL}(2;\mathbb{Z})_T \times \text{SL}(2;\mathbb{Z})_U  \to
     \begin{pmatrix}
        da'&-db'&-cb'&-ca'\\
        -dc'&dd'&cd'&cc'\\
        -bc'&bd'&ad'&ac'\\
        -ba'&bb'&ab'&aa'
    \end{pmatrix}  \in  \text{SO}^+(2,2;\mathbb{Z})\,.
\end{equation}

Consider now the shift vector $v$ given in \eqref{shiftvapp}. Under $\text{SL}(2;\mathbb{Z})_T\times \text{SL}(2;\mathbb{Z})_U$, it transforms as the $(\textbf{2},\textbf{2})$ representation. The vector $v$ represents the orbifold shift along the circle of radius $\mathcal{R}_5$, and induces a shift in the winding number $w_1$. Then, the \say{shifted} $T^2$ partition function will be given by\footnote{A more general discussion on shifted lattice sums can be found in e.g. \cite{Kiritsis:1997ca}.} 
\begin{equation}
    {Z}_{T^2}[k,l]=  \frac{1}{(\eta\widebar{\eta})^2}\sum_{\left\{n_i,w_i\right\} \in \mathbb{Z}^4}\,e^{\frac{2\pi i ln_1 }{p}} \,\widebar{q}^{\frac{1}{2}p_{\text{L}}^2(k)}\,q^{\frac{1}{2}p_{\text{R}}^2(k)}\,, \qquad i=1,2\,,
    \label{shiftedt2}
\end{equation}
where $k=0,\ldots,p-1$ labels the twisted sectors ($k=0$ corresponds to the untwisted sector) and summation over $l=0,\ldots,p-1$ implements the orbifold projection\footnote{If we denote the orbifold group element by $g$, with $g^p=1$, then the projection operator takes the form $P=\frac{1}{p}(1+g+g^2+\cdots + g^{p-1})$. }. Also,
\begin{equation}
    \begin{aligned}
        &p^2_{\text{L}}(k) = \frac{1}{2T_2U_2}|n_2-Un_1 +\widebar{T}\left(w_1+\tfrac{k}{p}\right) +\widebar{T}Uw_2|^2\,,\\
         &p^2_{\text{R}}(k) = \frac{1}{2T_2U_2}|n_2-Un_1 +{T}\left(w_1+\tfrac{k}{p}\right) +{T}Uw_2|^2\,.
    \end{aligned}
    \label{shifted momenta}
\end{equation}
Now, a generic $\text{Spin}(2,2;\mathbb{Z})$ T-duality transformation will  transform the shift vector $v$ to another shift vector $v'$, which will not leave the partition function invariant. This means that if we start with a model with shift vector $v$, after a T-duality transformation we will end up with an inequivalent model with shift vector $v'$. Hence, generic T-duality transformations will not be in the orbifold duality group.

However, as in the $\mathrm{SL}(2)$ example discussed before, there exists a subgroup of the T-duality group, $\text{Spin}(2,2;\mathbb{Z})$, that acts on the shift vector in such a way that the partition function remains invariant. This subgroup will be in the orbifold duality group. Firstly, a shift, $v\to v+P$, where $P$ is a lattice vector, corresponds to a shift  ${1}/{p}\to {1}/{p}+\mathbb{Z}$ in the phase factor in \eqref{shiftedt2}, and a shift of the momentum and winding numbers in \eqref{shifted momenta} by integers. Therefore, this is a symmetry of the partition function. In addition, a reflection, $v\to -v$ simply amounts to $l\to -l$ and $k\to -k$ in \eqref{shiftedt2} and \eqref{shifted momenta}, which is also a symmetry of the partition function. So, in order to determine the orbifold duality group associated with $T^2$, we need to find the subgroup of $\text{Spin}(2,2;\mathbb{Z})$ that generates these two symmetries (for a similar discussion see also \cite{Gregori:1997hi}, appendix C).

Let us first focus on the transformations that preserve the shift vector $v$ up to lattice translations. By combining \eqref{shiftvapp} with \eqref{sl2t} and \eqref{sl2u}, we find 
that the duality group
$\text{SL}(2;\mathbb{Z})_T$ is broken to $ \Gamma^1(p)_T$
and
$\text{SL}(2;\mathbb{Z})_U$ is broken to $ \Gamma_1(p)_U $, where
\begin{equation}
  \Gamma^1(p)_T = \Bigg\{ \begin{pmatrix}
       a &b\\
        c& d
    \end{pmatrix}\in  \text{SL}(2;\mathbb{Z})_T: a,d=1 \:\text{mod} \:\,p\,,\quad b=0\: \text{mod} \:\, p\Bigg\}\,,
    \label{gamma1t}
\end{equation}
\begin{equation}
 \quad\:\: \Gamma_1(p)_U = \Bigg\{ \begin{pmatrix}
       a' &b'\\
        c'& d'
    \end{pmatrix}\in  \text{SL}(2;\mathbb{Z})_U: a',d'=1 \:\text{mod} \:\,p\,,\quad c'=0\: \text{mod} \:\, p\Bigg\}\,.
    \label{gamma1u}
\end{equation}
The groups $\Gamma^1(p)$ and $\Gamma_1(p)$ are isomorphic (the isomorphism is simply given by transposition) but are embedded differently in $\text{SL}(2,\mathbb{Z})$.
The appearance of $\Gamma^1(p)_T$ instead of $\Gamma_1(p)_T$ is due to the embedding of $\text{SL}(2,\mathbb{Z})_T$ in $\text{Spin}(2,2,\mathbb{Z})$. It is important to stress that the surviving duality groups depend on the particular shift vector. For instance, a shift along the radius $\mathcal{R}_4$ that shifts $w_2$ instead of $w_1$, would lead to two identical $\Gamma^1(p)_T \times \Gamma^1(p)_U$
subgroups (see e.g. \cite{sen1995dual, ITOYAMA2022115667} or \cite{Gregori:1997hi} for examples).

Regarding the reflection, $v\to -v$, we can see that it can be realized by the values $a=d=-1,b=c=0$ in \eqref{sl2t}, and $a'=d'=-1, b'=c'=0$ in \eqref{sl2u}, which generate another two $\mathbb{Z}_2$ subgroups for $p>2$. Concluding, we have found that the T-duality group of $T^2$ is broken, due to the presence of the shift vector, to
\begin{equation}
\hat \Gamma^1(p)_T\times \hat \Gamma_1(p)_U \subset \text{SL}(2;\mathbb{Z})_T\times \text{SL}(2;\mathbb{Z})_U\,.
\end{equation}

Now, by combining all the above, we conclude that the T-duality group $\text{Spin}(6,6;\mathbb{Z})$ of $T^6$ is broken to the orbifold T-duality group
 \begin{equation}
\hat \Gamma^1(p)_T\times \hat \Gamma_1(p)_U \times
\mathcal{C}(\mathbb{Z})
\subset \text{SL}(2;\mathbb{Z})_T\times \text{SL}(2;\mathbb{Z})_U
\times \text{Spin}(4,4;\mathbb{Z}) \subset
\text{Spin}(6,6;\mathbb{Z}) \ .
  \label{Tsubg}
\end{equation}
For the supergravity theory, the shift breaks $\text{SL}(2;\mathbb{R})_T\times \text{SL}(2;\mathbb{R})_U$ to the subgroup of matrices
\begin{equation}
\Bigg\{ \begin{pmatrix}
       a &b\\
        c& d
    \end{pmatrix}\in  \text{SL}(2;\mathbb{R})_T, ~
    \begin{pmatrix}
       a' &b'\\
        c'& d'
    \end{pmatrix}\in  \text{SL}(2;\mathbb{R})_U : \quad
    da'=\pm 1 \:\text{mod} \:\,p\,,\quad 
b=0\: \text{mod} \:\, p, \quad c'=0\: \text{mod} \:\, p,
\Bigg\}\,,
    \label{Congr}
\end{equation}
which is the group
\begin{equation}
    (\mathbb{R}\times
\mathbb{Z}_2) \ltimes
\left(\hat  \Delta^1(p)\times  \hat  \Delta_1(p)\right) \,,
\end{equation}
where $\hat{\Delta}^1(p)$ is defined by
\begin{equation}
 \hat  \Delta^1(p) = \Bigg\{ \begin{pmatrix}
       a &b\\
        c& d
    \end{pmatrix}\in  \text{SL}(2;\mathbb{R}):\quad d=\pm 1 \:\text{mod} \:\,p\,,\quad b=0\: \text{mod} \:\, p\Bigg\}\, .
\end{equation}

Then in the supergravity theory, $\text{Spin}(6,6) $ is broken to the product of this group with $\mathcal{C}$.

We now turn to the breaking of the S-duality group $\text{SL}(2)_S$.
Under $\text{SL}(2)_S$,  the perturbative 4-vector $v$ transforms into a non-perturbative 4-vector of NS5-brane charges and KK monopole charges.
 Then the shift vector transforms as the {\bf 2} representation under $\text{SL}(2)_S$, so that it can be regarded as part of an 8-vector  transforming as the $(\textbf{4},\textbf{2})$ representation of ${\rm Spin}(2,2)\times \text{SL}(2)_S$. However, this ${\rm Spin}(2,2)$ factorises as ${\rm Spin}(2,2)=\text{SL}(2)_T\times \text{SL}(2)_U$, so that  ${\rm Spin}(2,2)\times \text{SL}(2)_S=\text{SL}(2)_T\times \text{SL}(2)_U\times \text{SL}(2)_S$
 and under this the shift vector transforms as a $(\textbf{2},\textbf{2},\textbf{2})$. 
 The non-zero component of (\ref{shiftvapp}) is then part of an $\text{SL}(2)_S$ doublet of the form (\ref{shiftvvapp}).  From the earlier discussion, the subgroup of
 $\text{SL}(2;\mathbb{Z})_S$ preserving this up to a sign and up to a lattice vector is $\hat \Gamma^1(p)_S$, so that the S-duality is broken to this subgroup.
Then
$\mathrm{SL}(2;\mathbb{Z})_T\times \mathrm{SL}(2;\mathbb{Z})_U\times \mathrm{SL}(2;\mathbb{Z})_S$ 
is broken to the subgroup $\hat \Gamma^1(p)_T \times
\hat \Gamma_1(p)_U\times
\hat \Gamma^1(p)_S$ and we have the final result that the U-duality group is broken to 
\begin{equation}\label{Uduality}
    \mathcal{K}(\mathbb{Z}) = \hat \Gamma^1(p)_T \times
\hat \Gamma_1(p)_U\times
\hat \Gamma^1(p)_S\times\mathcal{ C}(\mathbb{Z})\,.
\end{equation}

For the supergravity limit, the subgroup of $\mathrm{SL}(2)_S$ preserving the shift up to a sign and a lattice vector is $ \hat  \Delta^1(p)$. However, the subgroup of 
$\mathrm{SL}(2)_T\times \mathrm{SL}(2)_U\times \mathrm{SL}(2)_S$ preserving the shift is slightly larger than the product of the T-duality group given above.
 It is the group of matrices
 \begin{equation}
 \begin{aligned}
\Bigg\{ \begin{pmatrix}
       a &b\\
        c& d
    \end{pmatrix}\in  \text{SL}(2;\mathbb{R})_T, ~  &
    \begin{pmatrix}
       a' &b'\\
        c'& d'
    \end{pmatrix}\in  \text{SL}(2;\mathbb{R})_U 
    , ~
    \begin{pmatrix}
       a'' &b''\\
        c''& d''
    \end{pmatrix}\in  \text{SL}(2;\mathbb{R})_S
    :  \\
  &  da'd''=\pm 1 \:\text{mod} \:\,p\,,~ 
b=0\: \text{mod} \:\, p, ~ c'=0\: \text{mod} \:\, p,~b''=0\: \text{mod} \:\, p
\Bigg\}\,.
\end{aligned}
    \label{Congrs}
\end{equation}
The supergravity duality group is then the product of this with $\mathcal{C}$.

\section{An $STU$-like  $\mathbb{Z}_{6}$ model}
\label{z12 with nv=3 and nh=0}

In this section we discuss an $\mathcal{N}=2, D=4$ model, with three vector multiplets and two hypermultiplets arising in the massless untwisted orbifold sector. This model can be obtained from a circle reduction of an $\mathcal{N}=2, D=5$ freely acting $\mathbb{Z}_{6}$ orbifold model of IIB string theory on $T^5$ studied in \cite{Gkountoumis:2024dwc}. We consider a $\mathbb{Z}_{6}$ orbifold with the following mass parameters
\begin{equation}
    m_1={\pi}\ , \qquad m_2= \frac{\pi}{3}\ ,\qquad m_3=\frac{2\pi}{3} \ ,\qquad m_4=0\ .
\end{equation}
The corresponding twist vectors are (cf.  \eqref{relation twist vectors-mass parameters})
\begin{equation}
    \tilde{u}=\left(\frac{5}{6},\frac{1}{6}\right)\ ,\qquad u=\left(\frac{1}{6},\frac{1}{6}\right)\,,
    \label{twist vectors for primary example}
\end{equation}
which satisfy \eqref{modular conditions on twist vectors} (for this model $p=6$), and the appropriate lattice  is $\Gamma^{4,4}(A_2\oplus A_2)\oplus \Gamma^{2,2}$. It is easy to verify that \eqref{modcond2} is also satisfied, as
\begin{equation}
    p_{\text{L}}\mathcal{N}_{\text{L}}^{\,3}p_{\text{L}} - p_{\text{R}}\mathcal{N}_{\text{R}}^{\,3}p_{\text{R}}=-(p_{\text{L}}^2-p_{\text{R}}^2) \in 2\mathbb{Z}\ .
\end{equation}
As a result, the theory is modular invariant. Notice that  $\Gamma^{2,2}$ is left invariant under the twist. We picked this model because it has only NS-NS massless scalars and vectors. Consequently, the classical moduli space can be determined by simply computing the subgroup of the T-duality group that preserves the orbifold twist and shift, as explained in section \ref{duality groups of orbifolded theory}.

In the following, we will discuss the massless spectrum of our model in the untwisted sector, and the spectrum of the lightest states in the twisted sectors. Due to the orbifold shift, states in the twisted sectors are generically massive. However, there exist special lines in the bulk of the moduli space in which a finite number of massive twisted states become massless, which will be discussed in \autoref{Massless states at finite distance}.  Also, there exist points at infinite distance, where infinite towers of states become massless, and we will analyse these in \autoref{Massless states at infinite distance}.

\subsection{Untwisted sector}

The massless spectrum in the untwisted sector of our $\mathbb{Z}_6$ model can be obtained by circle reduction of a five-dimensional model, with the same mass parameters and twist vectors, that was studied in \cite{Gkountoumis:2024dwc}. The untwisted massless spectrum of the five-dimensional model consists of the $\mathcal{N}=2$ gravity multiplet, two vector multiplets, and two hypermultiplets, and all the fields come from the NS-NS and NS-R sectors. On reduction on a circle we simply get one additional vector multiplet. Then the massless untwisted orbifold spectrum consists of the  $\mathcal{N}=2$, $D=4$ gravity multiplet coupled to 3 vector multiplets and 2 hypermultiplets; this can also be seen from the detailed analysis of the spectrum below.

The untwisted orbifold spectrum can be obtained from the partition function, which can be expanded as \cite{Gkountoumis:2023fym}
\begin{equation}
    \begin{aligned}
    {Z}[0,l]=(q\widebar{q})^{-\frac{1}{2}}\sum_{\left\{n_i,w_i\right\} \in \mathbb{Z}^4}\,e^{\frac{2\pi i ln_1 }{p}} \,\widebar{q}^{\frac{1}{2}p_{\text{L}}^2(0)}\,q^{\frac{1}{2}p_{\text{R}}^2(0)}\, 
    \sum_{r,\tilde{r}} q^{\frac{1}{2}{r}^2}\, (\widebar{q})^{\frac{1}{2}\tilde{r}^2}\,e^{2\pi il (\tilde{r}\cdot \tilde{u}-r\cdot u)}\, \left(1+\cdots\right)\,.
    \end{aligned}
    \label{generic partition untwisted}
\end{equation}
Here, $r=(r_1,r_2,r_3,r_4)$ (and similarly $\tilde{r}$) is an SO(8) weight vector with each component ${r}_i\in \mathbb{Z}$ in the NS-sector and ${r}\in \mathbb{Z}+\tfrac{1}{2}$ in the R-sector, while the  left and right-moving momenta, $p^2_{\text{L/R}}(k)$ are given in \eqref{shifted momenta}. The GSO projection is $\sum_{i=1}^4 r_i \in 2\mathbb{Z}+1$ in the NS-sector and $\sum_{i=1}^4 r_i \in 2\mathbb{Z}$ in the R-sector. 
Finally, the dots denote contributions to the partition function from higher excited bosonic oscillator states. 

Now, string states are constructed by tensoring left and right-movers, and a generic state is characterised by the product $\tilde{r}\times r$. In order to implement the orbifold projection we need to sum over $l$ and divide by the orbifold rank $p$. Then, the degeneracy of a state in the untwisted sector will be given by
\begin{equation}
    D(k=0)=\frac{1}{p}\sum_{l=0}^{p-1}e^{2\pi il \left[(\tilde{r}\cdot \tilde{u}-r\cdot u)+\frac{n_1}{p}\right]}\ .
    \label{degeneracy untwisted}
\end{equation}
Note that states with trivial orbifold charge, i.e.\ trivial phase $e^{2\pi il [(\tilde{r}\cdot \tilde{u}-r\cdot u)]}$, will survive the orbifold projection and will have degeneracy 1. States with non-trivial orbifold charge will survive the orbifold projection with the addition of appropriate momentum number $n_1$, such that they will become massive and will come with degeneracy 1. 

The masses of left and right-moving states can be read off from the exponents of $\widebar{q}$ and $q$, respectively. In particular, the mass formulae read
\begin{equation}
\begin{aligned}
    &\frac{\alpha'm^2_{\text{L}}(0)}{2}=\frac{1}{2}\tilde{r}^2+\frac{1}{2}p^2_{\text{L}}(0)-\frac{1}{2}+\widetilde{N}\,,\\
    &\frac{\alpha'm^2_{\text{R}}(0)}{2}=\frac{1}{2}{r}^2+\frac{1}{2}p^2_{\text{R}}(0)-\frac{1}{2}+N\,.
\end{aligned}
 \label{untwisted masses}
 \end{equation}
Here, $\widetilde{N}$ and $N$ are integers, which refer to the bosonic occupation number of higher excited left and right-moving oscillator states, respectively. In the untwisted sector, the lightest states satisfy $\widetilde{N}=N=0$. 

Let us now move on to the construction of massless states in the untwisted sector. First, we list in \autoref{untwisted stu states} the NS and R-sector $\text{SO}(8)$ weight vectors for the lightest left and right-moving states that survive the GSO projection; all of these are massless in the absence of momentum and/or winding modes. 
\renewcommand{\arraystretch}{2}
\begin{table}[h!]
\centering
 \begin{tabular}{|c|c|}
    \hline
    Sector & SO(8) weight  \\
    \hline
    \hline
  NS & $(\pm1,0,0,0)$,\: $(0,\underline{0,0,\pm1})$  \\
  \hline
   \multirow{2}{*}{R}  & $\pm(\frac{1}{2},\frac{1}{2},\frac{1}{2},\frac{1}{2})$,\: $\pm(-\frac{1}{2},-\frac{1}{2},\frac{1}{2},\frac{1}{2})$ \\
   \cline{2-2}
   & $\pm(\frac{1}{2},-\frac{1}{2},\frac{1}{2},-\frac{1}{2})$,\: $\pm(-\frac{1}{2},\frac{1}{2},\frac{1}{2},-\frac{1}{2})$ \\
   \hline
    \end{tabular}
\captionsetup{width=.9\linewidth}
\caption{Here we list the weight vectors of the lightest left and right-moving states in the untwisted sector. Underlying denotes permutation; e.g. $(0,\underline{0,0,1})$ denotes the states $(0,1,0,0)$, $(0,0,1,0)$ and $(0,0,0,1)$.}
\label{untwisted stu states}
\end{table}
\renewcommand{\arraystretch}{1}

Massless string states in the untwisted sector must have a trivial orbifold charge, i.e.\ they must be invariant under $\mathbb{Z}_6$, and can be classified based on their helicity in $4D$. The helicity of a state $(\tilde{r}_1,\tilde{r}_2,\tilde{r}_3,\tilde{r}_4)\times (r_1,r_2,r_3,r_4)$ is equal to $\tilde{r}_1-r_1$ \cite{font2005introduction}. As we can see from \eqref{twist vectors for primary example} and \eqref{degeneracy untwisted}, massless states should satisfy $5\tilde{r}_3+\tilde{r}_4-(r_3+r_4)=0$ mod 6, which is the case only in the NS-NS and NS-R sectors. In the NS-NS sector we find the following massless states
\begin{equation}
\begin{aligned}\label{untwisted_massless}
   & (\pm1,0,0,0) \times (\pm1,0,0,0) : (\pm2) + 2\times (0)\,,\\
    & (\pm1,0,0,0) \times (0,\pm1,0,0) :2\times (\pm1)\,,\\
    &(0,\pm1,0,0) \times (\pm1,0,0,0) : 2\times (\pm1)\,,\\
    &(0,\pm1,0,0) \times (0,\pm1,0,0) : 4\times (0)\,,\\
    \pm&[(0,0,1,0) \times (0,0,\underline{-1,0})]: 4\times (0)\,,\\
   \pm &[(0,0,0,1) \times (0,0,\underline{1,0})]: 4\times (0)\,.
    \end{aligned}
\end{equation}
In total, we find the helicities that correspond to the graviton, $(\pm2)$, 4 massless vectors, $4\times(\pm1)$, and 14 scalars, $14\times(0)$. The massless states in the NS-R sector can be constructed in a similar way. We find 2 gravitini, $2\times(\pm\tfrac{3}{2})$, and 10 dilatini, $10\times(\pm\tfrac{1}{2})$. Together the fields from the NS-NS and NS-R sectors form the $\mathcal{N}=2$, $D=4$ gravity multiplet, 3 vector multiplets and 2 hypermultiplets. 

Note that the scalars in the 3 vector multiplets are the complex $S,\,T$ and $U$ moduli, where $T$ and $U$ are the $T^2$ moduli, defined in \eqref{T,Udefap}, and $S$ is defined by
\begin{equation}
    S = a + i e^{-2\phi_4}\,,
\end{equation}
Here $a$ is a scalar that is dual to the NS-NS $B$-field in four dimensions, and is usually referred to as the axion, and $\phi_4$ is a scalar parametrising the four-dimensional string coupling by $\lambda_4=\langle e^{\phi_4}\rangle$. There are also 4 complex scalars in the two hypermultiplets.

Of course, all the aforementioned states in the NS-NS and NS-R sectors come with infinite towers of momentum and winding modes along $T^2$, and their masses are given by (cf. \eqref{momentauntwisted} and \eqref{untwisted masses})
\begin{equation}
\begin{aligned}
    &\frac{\alpha'm^2_{\text{L}}(0)}{2}=\frac{1}{2}p^2_{\text{L}}(0)=\frac{1}{4T_2U_2}|n_2-Un_1 +\widebar{T}w_1 +\widebar{T}Uw_2|^2\,,\\
    &\frac{\alpha'm^2_{\text{R}}(0)}{2}=\frac{1}{2}p^2_{\text{R}}(0)=\frac{1}{4T_2U_2}|n_2-Un_1 +{T}w_1 +{T}Uw_2|^2\,.
    \label{toweruntwisted}
\end{aligned}
 \end{equation}
Note that in these formulae the momentum number $n_1$ should obey $n_1=0$ mod $6$, such that the states are invariant under the orbifold symmetry.  Physical states should also satisfy the level-matching condition, $m^2_{\text{L}}=m^2_{\text{R}}$, which in this case reads
\begin{equation}
    n_1w_1+n_2w_2=0\,.
\end{equation}

\subsection{Twisted sectors}
As in the untwisted sector, the spectrum of lightest states in each twisted sector can be obtained from the partition function, which can be expanded as \cite{Gkountoumis:2023fym}
\begin{equation}
    \begin{aligned}
    {Z}[k,l]=&{\chi} [k,l]\,\tilde{\chi}[k,l](q\widebar{q})^{-\frac{1}{2}}\,e^{i(\tilde{\varphi}-\varphi)}q^{E_k}\,(\widebar{q})^{\widetilde{E}_k}\sum_{\left\{n_i,w_i\right\} \in \mathbb{Z}^4}\,e^{\frac{2\pi i ln_1 }{p}} \,\widebar{q}^{\frac{1}{2}p_{\text{L}}^2(k)}\,q^{\frac{1}{2}p_{\text{R}}^2(k)}\, \times\\
    & \sum_{r,\tilde{r}} q^{\frac{1}{2}({r}+k u)^2}\, (\widebar{q})^{\frac{1}{2}(\tilde{r}+k \tilde{u})^2}\,e^{2\pi il (\tilde{r}\cdot \tilde{u}-r\cdot u)}\, e^{2\pi il k (\tilde{u}^2-u^2)}\, \left(1+\cdots\right)\,.
    \end{aligned}
    \label{generic partition twisted}
\end{equation}
Here, as in the untwisted sector, the dots denote contributions to the partition function from higher excited bosonic oscillator states, while
\begin{equation}
   \chi[k,l]= \prod^4_{i=3}2 \sin(\pi \text{\footnotesize{gcd}}(k,l) u_i)\ ,
   \label{fixed points}
\end{equation}
is the number of \say{chiral} fixed points\footnote{The orbifolds that we consider have fixed points on $T^4$. However, due to the shift along the one circle of $T^2$, there are no points left invariant under the full orbifold action. Also, $\text{\footnotesize{gcd}}(k,l)$ denotes the greatest common divisor of $k,l$ with the convention $\text{\footnotesize{gcd}}(a,0)=\text{\footnotesize{gcd}}(0,a)=a$.}. We note here that equation \eqref{fixed points} is valid for $ku_{3,4} \notin \mathbb{Z}$. If there exists $j\in [3,4]$ such that $ku_j \in \mathbb{Z}$, $ \chi[k,l]$ should be divided by $2\sin(\pi l u_j)$ for $l\neq 0$, and replaced by $\prod_{i\neq j,ku_i \notin \mathbb{Z}}2 \sin(\pi k u_i)$ for $l=0$  (see \cite{katsuki1990zn} for a relevant discussion). In addition, from the expansion of the bosonic piece of the partition function we obtain the phase factor
\begin{equation}
    \varphi= 2\pi  \sum_{u_i\notin \mathbb{Z}}\left(\frac{1}{2}-k u_i\right)l u_i\,,
    \label{shift phase factor}
\end{equation}
and a shift to the zero point energy given by
\begin{equation}
    E_k = \sum_{u_i\notin \mathbb{Z}}\frac{1}{2}ku_i (1-ku_i)\,.
    \label{shifted energy}
\end{equation}
Note that if $ku_i>1$, we should substitute $ku_i\to ku_i-1$ in \eqref{shifted energy}, so that we actually compute the energy of the lowest order terms. This leads to the same modification of the phase factor in \eqref{shift phase factor}, together with an overall shift $\varphi\to \varphi +\pi$. Regarding the expressions for $\tilde{\chi}[k,l],\tilde{\varphi}$ and $\widetilde{E}_k$, these can be simply obtained from \eqref{fixed points}-\eqref{shifted energy} by substituting $u\to\tilde{u}$. The expressions for the left and right-moving momenta, $p^2_{\text{L/R}}(k)$, are given in \eqref{shifted momenta}. We mention here that in the formulae \eqref{fixed points}-\eqref{shifted energy} we consider $u_i>0$, as sending $u_i\to-u_i$, for $i=3$ and/or $4$, leaves the bosonic piece of the partition function invariant.

Finally, in order to implement the orbifold projection, we fix $k$, sum over $l$ and divide by the orbifold rank. Then, the degeneracy of a state with no bosonic oscillator excitations in the $k$-th twisted sector is given by (see also \cite{ibanez1988heterotic,font1990construction} for a relevant discussion)
\begin{equation}
   D(k)= \frac{1}{p}\sum_{l=0}^{p-1}{\chi} [k,l]\,\tilde{\chi}[k,l] e^{2\pi il \left[(\tilde{r}\cdot \tilde{u}-r\cdot u)+k(\tilde{u}^2-u^2)+\frac{n_1}{p}\right]+i(\tilde{\varphi}-\varphi)}\,.
   \label{degeneracy twisted}
\end{equation}
In the twisted sectors the mass formulae read
\begin{equation}
\begin{aligned}
    &\frac{\alpha'm^2_{\text{L}}(k)}{2}=\frac{1}{2}(\tilde{r}+k\tilde{u})^2+\frac{1}{2}p^2_{\text{L}}(k)+\widetilde{E}_k-\frac{1}{2}+\widetilde{N}\,,\\
    &\frac{\alpha'm^2_{\text{R}}(k)}{2}=\frac{1}{2}({r}+k{u})^2+\frac{1}{2}p^2_{\text{R}}(k)+{E}_k-\frac{1}{2}+N\,,
\end{aligned}
 \label{twisted masses}
 \end{equation}
where $p_{\text{L}}^2,\,p_{\text{R}}^2$ are defined in \eqref{shifted momenta}. As in the untwisted sector, $\widetilde{N}$ and $N$ refer to the bosonic occupation number of higher excited left and right-moving oscillator states, respectively. However, in the twisted sectors $\widetilde{N}$ and $N$ are not integers because the twisted boundary conditions of the $T^4$ coordinates \eqref{boundaryconditions} result in a shift of the moding of the corresponding oscillators. In particular, the modings of the bosonic oscillators along the torus directions read (see e.g. \cite{font2005introduction}) 
\begin{equation}
\begin{aligned}
    &\tilde{\alpha}^1_{n-k\tilde{u}_3}\,, \quad \bar{\tilde{\alpha}}^1_{n+k\tilde{u}_3}\,, \quad{\alpha}^1_{n+k{u}_3}\,,\quad \bar{\alpha}^1_{n-k{u}_3}\,,\qquad n\in\mathbb{Z}\,,\\
    &\tilde{\alpha}^2_{n-k\tilde{u}_4}\,, \quad \bar{\tilde{\alpha}}^2_{n+k\tilde{u}_4}\,, \quad{\alpha}^2_{n+k{u}_4}\,,\quad \bar{\alpha}^2_{n-k{u}_4}\,,\qquad n\in\mathbb{Z}\,.
    \end{aligned}
    \label{shifted oscillators}
\end{equation}
Here, the left-movers are denoted by a tilde, and a bar denotes the complex conjugate. As an example, consider a generic state in a $k$ twisted sector denoted by $\ket{\tilde{r},r}_k$. We can act on this state with a left-moving creation operator, i.e. $\tilde{\alpha}^i_{-k\tilde{u}_i}\ket{\tilde{r},r}_k$. Then, $\widetilde{N}= k\tilde{u}_i$, and the degeneracy of the state \eqref{degeneracy twisted} is modified by the addition of a phase $e^{2\pi i l\tilde{u}_i}$ (see e.g. \cite{font2005introduction}).

\subsubsection{$k=1$ sector}
Let us now work out the spectrum of lightest states in the $k=1$ sector of our orbifold model. We mention here that, in general, the spectrum in an orbifold $k$-twisted sector is identical with the spectrum of the $(p-k)$-twisted sector. This is due to the fact that the partition function of the $k$-twisted sector is equal to the partition function of the $(p-k)$-twisted sector. As a result, the spectrum in the $k=1$ sector is the same as in the $k=5$ sector.

First, we compute the shift in the zero point energies (cf. \eqref{twist vectors for primary example}, \eqref{shifted energy})
\begin{equation}
    \widetilde{E}_1 =E_1=\frac{5}{36}\,,
\end{equation}
together with the degeneracy  of a state characterised by generic weight vectors $\tilde{r}$ and $r$ and without bosonic oscillator excitations (cf. \eqref{twist vectors for primary example}, \eqref{fixed points}, \eqref{shift phase factor}, \eqref{degeneracy twisted}), which is given by
\begin{equation}
    D(1)=\frac{1}{6}\sum_{l=0}^{5} e^{\frac{2\pi il}{6}[5\tilde{r}_3+\tilde{r}_4-(r_3+r_4) +2 +n_1] }\,.
    \label{degeneracy1}
\end{equation}
(We use the notation $D(k)$ for the degeneracy in the
$k$'th twisted sector.)
We list the weight vectors for the lightest left and right-moving states in the $k=1$ sector in \autoref{k=1 stu states}.
\renewcommand{\arraystretch}{2}
\begin{table}[h!]
\centering
 \begin{tabular}{|c|c|c|}
    \hline
    Sector &  $\tilde{r}$ &   ${r}$ \\
    \hline
    \hline
  NS & $(0,0,-1,0)$ &$(0,0,\underline{0,-1})$ \\
  \hline
   \multirow{2}{*}{R}  & $(\frac{1}{2},\frac{1}{2},-\frac{1}{2},-\frac{1}{2})$& $(\frac{1}{2},\frac{1}{2},-\frac{1}{2},-\frac{1}{2})$ \\
   \cline{2-3}
   & $(-\frac{1}{2},-\frac{1}{2},-\frac{1}{2},-\frac{1}{2})$& $(-\frac{1}{2},-\frac{1}{2},-\frac{1}{2},-\frac{1}{2})$ \\
   \hline
    \end{tabular}
\captionsetup{width=.9\linewidth}
\caption{The weight vectors of the lightest left and right-moving states in the $k=1$ twisted sector.}
\label{k=1 stu states}
\end{table}
\renewcommand{\arraystretch}{1}

Let us start with the construction of states in the R-R sector, in which we have 
\begin{equation}
\begin{aligned}
    & (\tfrac{1}{2},\tfrac{1}{2},-\tfrac{1}{2},-\tfrac{1}{2})\times (\tfrac{1}{2},\tfrac{1}{2},-\tfrac{1}{2},-\tfrac{1}{2}) : (0)\,,\\
    & (\tfrac{1}{2},\tfrac{1}{2},-\tfrac{1}{2},-\tfrac{1}{2})\times (-\tfrac{1}{2},-\tfrac{1}{2},-\tfrac{1}{2},-\tfrac{1}{2}) : (1)\,,\\
    & (-\tfrac{1}{2},-\tfrac{1}{2},-\tfrac{1}{2},-\tfrac{1}{2})\times (\tfrac{1}{2},\tfrac{1}{2},-\tfrac{1}{2},-\tfrac{1}{2}) : (-1)\,,\\
     &(-\tfrac{1}{2},-\tfrac{1}{2},-\tfrac{1}{2},-\tfrac{1}{2})\times (-\tfrac{1}{2},-\tfrac{1}{2},-\tfrac{1}{2},-\tfrac{1}{2}) : (0)\,.
\end{aligned}
\end{equation}
Here we denote a state of helicity $h$ by $(h)$.
Using \eqref{degeneracy1}, we can see that the above states are orbifold invariant for $n_1=0$ mod $6$, and have degeneracy $1$. From the spacetime point of view, we have the helicities that correspond to 1 massive vector $(\pm 1,0)$ and 1 scalar $(0)$. Also, from the mass formulae \eqref{twisted masses} we get (for $\widetilde{N}=N=0$)
\begin{equation}
\begin{aligned}
    &\frac{\alpha'm^2_{\text{L}}(1)}{2}=\frac{1}{2}p^2_{\text{L}}(1)=\frac{1}{4T_2U_2}\left|n_2-Un_1 +\widebar{T}\left(w_1+\tfrac{1}{6}\right) +\widebar{T}Uw_2\right|^2\,,\\
    &\frac{\alpha'm^2_{\text{R}}(1)}{2}=\frac{1}{2}p^2_{\text{R}}(1)=\frac{1}{4T_2U_2}\left|n_2-Un_1 +{T}\left(w_1+\tfrac{1}{6}\right) +{T}Uw_2\right|^2\,,
\end{aligned}
    \label{vmk1}
\end{equation}
where now $n_1=0$ mod $6$. Of course, physical states should also satisfy level-matching, that is $m^2_{\text{L}}(1)=m^2_{\text{R}}(1)$, which yields
\begin{equation}
    n_1\left(w_1+\frac{1}{6}\right) + n_2w_2 = 0\,.
    \label{levelvmk1}
\end{equation}
In the R-NS sector we find the following states
\begin{equation}
    \begin{aligned}
       & (\tfrac{1}{2},\tfrac{1}{2},-\tfrac{1}{2},-\tfrac{1}{2})\times (0,0,\underline{0,-1}) : 2\times (\tfrac{1}{2})\,,\\
    & (-\tfrac{1}{2},-\tfrac{1}{2},-\tfrac{1}{2},-\tfrac{1}{2})\times(0,0,\underline{0,-1}) : 2\times (-\tfrac{1}{2})\,. 
    \end{aligned}
\end{equation}
Similarly with the R-R sector, these states are orbifold invariant for $n_1=0$ mod $6$, and have degeneracy $1$. They  correspond to 2 dilatini with helicity $(\pm \tfrac{1}{2})$ and mass given in \eqref{vmk1}. The fields from the R-R and R-NS sectors form 1 tower of $4D$ massive vector multiplets, with mass \eqref{vmk1}.

Let us now move on to the NS-NS sector, in which we have
\begin{equation}
    (0,0,-1,0)\times (0,0,\underline{0,-1}) : 2\times (0)\,.
    \label{NSNSk=1}
\end{equation}
From \eqref{degeneracy1}, we can see that the above states are orbifold invariant for $n_1=2$ mod $6$, and have degeneracy $1$. So, in the NS-NS sector we find 2 scalars, denoted by $2\times (0)$. From the mass formulae \eqref{twisted masses} we get (for $\widetilde{N}=N=0$)
\begin{equation}
\begin{aligned}
    &\frac{\alpha'm^2_{\text{L}}(1)}{2}=\frac{1}{2}p^2_{\text{L}}(1)-\frac{1}{3}=\frac{1}{4T_2U_2}|n_2-Un_1 +\widebar{T}(w_1+\tfrac{1}{6}) +\widebar{T}Uw_2|^2-\frac{1}{3}\,,\\
    &\frac{\alpha'm^2_{\text{R}}(1)}{2}=\frac{1}{2}p^2_{\text{R}}(1)=\frac{1}{4T_2U_2}|n_2-Un_1 +{T}(w_1+\tfrac{1}{6}) +{T}Uw_2|^2\,,
\end{aligned}
\label{hmk1}
\end{equation}
and the level-matching condition becomes
\begin{equation}
    n_1\left(w_1+\frac{1}{6}\right) + n_2w_2 = \frac{1}{3}\,,
    \label{levelhmk1}
\end{equation}
which gives $n_1=2$ mod $6$.

In the NS-R sector we find the following states
\begin{equation}
    \begin{aligned}
  &   (0,0,-1,0)\times (\tfrac{1}{2},\tfrac{1}{2},-\tfrac{1}{2},-\tfrac{1}{2}) :  (-\tfrac{1}{2})\,,\\
  &  (0,0,-1,0)\times (-\tfrac{1}{2},-\tfrac{1}{2},-\tfrac{1}{2},-\tfrac{1}{2}) :  (\tfrac{1}{2})\,.
    \end{aligned}
    \label{NSRk=1}
\end{equation}
As in the NS-NS sector, these states are orbifold invariant for $n_1=2$ mod $6$, and have degeneracy $1$. They correspond to 1 dilatino $(\pm\tfrac{1}{2})$ with mass \eqref{hmk1}. The fields from the NS-NS and NS-R sectors in the $k=1$ sector, constitute half of the field content of a hypermultiplet\footnote{The other half arises from the $k=5$ twisted sector, which, as we have already mentioned, is equivalent to the $k=1$ sector.}. 

Now, we can also act on the NS-NS states \eqref{NSNSk=1} with the left-moving creation operators $\tilde{a}^2_{-{1}/{6}}$ or $\widebar{\tilde{a}}^1_{-{1}/{6}}$ (cf. \eqref{shifted oscillators}, \eqref{twist vectors for primary example}). These will contribute to $m_{\text{L}}^2(1)$ in \eqref{hmk1} by a factor of $\widetilde{N}=1/6$, and to the degeneracy of those states \eqref{degeneracy1} by a factor $e^{2\pi i l/6}$. So, these states will be orbifold invariant for $n_1=1$ mod $6$. Putting everything together, we find 4 scalars with mass
\begin{equation}
\begin{aligned}
    &\frac{\alpha'm^2_{\text{L}}(1)}{2}=\frac{1}{2}p^2_{\text{L}}(1)-\frac{1}{3}+\frac{1}{6}=\frac{1}{4T_2U_2}|n_2-Un_1 +\widebar{T}(w_1+\tfrac{1}{6}) +\widebar{T}Uw_2|^2-\frac{1}{6}\,,\\
    &\frac{\alpha'm^2_{\text{R}}(1)}{2}=\frac{1}{2}p^2_{\text{R}}(1)=\frac{1}{4T_2U_2}|n_2-Un_1 +{T}(w_1+\tfrac{1}{6}) +{T}Uw_2|^2\,,
\end{aligned}
\label{hmk1-case2}
\end{equation}
and the level-matching condition becomes
\begin{equation}
    n_1\left(w_1+\frac{1}{6}\right) + n_2w_2 = \frac{1}{6}\,,
    \label{levelhmk1-case2}
\end{equation}
with $n_1=1$ mod $6$. The above discussion applies also to the NS-R states \eqref{NSRk=1}, where we find 2 dilatini with mass \eqref{hmk1-case2}. So, in the $k=1$ sector we also find 1 tower of massive hypermultiplets with mass \eqref{hmk1-case2}.

Finally, we can act on the NS-NS states \eqref{NSNSk=1} with combinations of two left-moving creation operators $\tilde{a}^2_{-{1}/{6}}$ and/or $\widebar{\tilde{a}}^1_{-{1}/{6}}$ (cf. \eqref{shifted oscillators}, \eqref{twist vectors for primary example}). These will contribute to $\alpha'm^2(1)_{\text{L}}/2$ in \eqref{hmk1} by a factor of $\widetilde{N}=1/3$, and to the degeneracy of those states \eqref{degeneracy1} by a factor $e^{2\pi i l/3}$. So, these states will be orbifold invariant for $n_1=0$ mod $6$. Putting everything together, we find 6 scalars with mass
\begin{equation}
\begin{aligned}
    &\frac{\alpha'm^2_{\text{L}}(1)}{2}=\frac{1}{2}p^2_{\text{L}}(1)-\frac{1}{3}+\frac{1}{3}=\frac{1}{4T_2U_2}|n_2-Un_1 +\widebar{T}(w_1+\tfrac{1}{6}) +\widebar{T}Uw_2|^2\,,\\
    &\frac{\alpha'm^2_{\text{R}}(1)}{2}=\frac{1}{2}p^2_{\text{R}}(1)=\frac{1}{4T_2U_2}|n_2-Un_1 +{T}(w_1+\tfrac{1}{6}) +{T}Uw_2|^2\,,
\end{aligned}
\label{hmk1-case3}
\end{equation}
and the level-matching condition becomes
\begin{equation}
    n_1\left(w_1+\frac{1}{6}\right) + n_2w_2 =0\,,
    \label{levelhmk1-case3}
\end{equation}
with $n_1=0$ mod $6$. Similarly, we can act with the left-moving oscillators $\tilde{a}^2_{-{1}/{6}}$ and $\widebar{\tilde{a}}^1_{-{1}/{6}}$ on the NS-R states \eqref{NSRk=1}. Then we find 3 dilatini with mass \eqref{hmk1-case3}. In this case, the fields from the NS-NS and NS-R sectors constitute the field content of 1 and a half hypermultiplet\footnote{The other 1 and a half hypermultiplet arises from the $k=5$ twisted sector.}.

\subsubsection{$k=2$ sector}
Now we move on to the construction of states in the $k=2$ sector, which is equivalent to the $k=4$ sector. The shift in the zero point energies is (cf. \eqref{shifted energy}) 
\begin{equation}
    \widetilde{E}_2=E_2=\frac{2}{9}\,,
\end{equation}
and the degeneracy of a generic state is given by\footnote{Here, we have introduced an additional phase factor $e^{\pi il}$, in order to account for an extra minus sign arising from the fixed points.} (cf. \eqref{fixed points}, \eqref{shift phase factor}, \eqref{degeneracy twisted})
\begin{equation}
    D(2)= \frac{1}{6}\sum_{l=0}^{5}{\chi} [2,l]\,\tilde{\chi}[2,l] e^{\frac{2\pi il}{6}[5\tilde{r}_3+\tilde{r}_4-(r_3+r_4) -2 +n_1] }\,,
    \label{deg2}
\end{equation}
where
\begin{equation}
   {\chi} [2,l]\, \tilde{\chi}[2,l]= 
\begin{dcases}
   \: 9& \text{for}\quad l=0,2,4\,.\\
   \: 1& \text{for} \quad l=1,3,5\,.
\end{dcases}
\label{chis2}
\end{equation}
We list the weight vectors for the lightest left and right-moving states of the $k=2$ sector in \autoref{k=2 stu states}.
\renewcommand{\arraystretch}{2}
\begin{table}[h!]
\centering
 \begin{tabular}{|c|c|c|}
    \hline
    Sector &  $\tilde{r}$ &   ${r}$ \\
    \hline
    \hline
  \multirow{2}{*}{NS} & $(0,0,-1,0)$ &$(0,0,-1,0)$ \\
  \cline{2-3}
  & $(0,0,-2,-1)$ &$(0,0,0,-1)$ \\
  \hline
   \multirow{2}{*}{R}  & $(\frac{1}{2},-\frac{1}{2},-\frac{3}{2},-\frac{1}{2})$& $(\frac{1}{2},\frac{1}{2},-\frac{1}{2},-\frac{1}{2})$ \\
   \cline{2-3}
   & $(-\frac{1}{2},\frac{1}{2},-\frac{3}{2},-\frac{1}{2})$& $(-\frac{1}{2},-\frac{1}{2},-\frac{1}{2},-\frac{1}{2})$ \\
   \hline
    \end{tabular}
\captionsetup{width=.9\linewidth}
\caption{The weight vectors of the lightest left and right-moving states in the $k=2$ twisted sector.}
\label{k=2 stu states}
\end{table}
\renewcommand{\arraystretch}{1}

Now, we take tensor products between left and right-movers from \autoref{k=2 stu states}. In the NS-NS sector we find 
\begin{equation}
    \begin{aligned}
      &(0,0,-1,0) \times  (0,0,\underline{0,-1}) : 2\times (0)\,\\
       &(0,0,-2,-1) \times  (0,0,\underline{0,-1}) : 2\times (0)\,.
    \end{aligned}
    \label{k=2 NSNS}
\end{equation}
From \eqref{deg2} and \eqref{chis2}, we see that these states survive the orbifold projection for $n_1=0$ mod $6$, and have degeneracy 5. So, we find 20 scalars (0) with mass (cf. \eqref{twisted masses}, with $\widetilde{N}=N=0$)
\begin{equation}
\begin{aligned}
    &\frac{\alpha'm^2_{\text{L}}(2)}{2}=\frac{1}{2}p^2_{\text{L}}(2)=\frac{1}{4T_2U_2}|n_2-Un_1 +\widebar{T}(w_1+\tfrac{1}{3}) +\widebar{T}Uw_2|^2\,,\\
    &\frac{\alpha'm^2_{\text{R}}(2)}{2}=\frac{1}{2}p^2_{\text{R}}(2)=\frac{1}{4T_2U_2}|n_2-Un_1 +{T}(w_1+\tfrac{1}{3}) +{T}Uw_2|^2\,,
\end{aligned}
    \label{hmk2}
\end{equation}
where $n_1=0$ mod $6$, and the level-matching condition reads
\begin{equation}
     n_1\left(w_1+\frac{1}{3}\right) + n_2w_2 = 0\,.
     \label{levelmatch2}
\end{equation}
Note that the states in \eqref{k=2 NSNS}, can also survive the orbifold projection for $n_1=3$ mod $6$, and, in this case, they come with degeneracy $4$. The construction of the lightest states in the NS-R sector is similar with the NS-NS sector, so we omit the details. We find 10 dilatini $(\pm\tfrac{1}{2})$ with mass \eqref{hmk2}. So, from the NS-NS and NS-R sectors we obtain 5 towers of hypermultiplets with mass \eqref{hmk2}, where $n_1=0$ mod $6$, and 4 towers of hypermultiplets with mass \eqref{hmk2}, where $n_1=3$ mod $6$.

Let us now discuss the spectrum in the R-R sector, in which we have
\begin{equation}
  \begin{aligned}
      &(\tfrac{1}{2},-\tfrac{1}{2},-\tfrac{3}{2},-\tfrac{1}{2})\times (\tfrac{1}{2},\tfrac{1}{2},-\tfrac{1}{2},-\tfrac{1}{2}): (0)\,,\\
      &(\tfrac{1}{2},-\tfrac{1}{2},-\tfrac{3}{2},-\tfrac{1}{2})\times (-\tfrac{1}{2},-\tfrac{1}{2},-\tfrac{1}{2},-\tfrac{1}{2}): (1)\,,\\
      &(-\tfrac{1}{2},\tfrac{1}{2},-\tfrac{3}{2},-\tfrac{1}{2})\times (\tfrac{1}{2},\tfrac{1}{2},-\tfrac{1}{2},-\tfrac{1}{2}): (-1)\,,\\
      &(-\tfrac{1}{2},\tfrac{1}{2},-\tfrac{3}{2},-\tfrac{1}{2})\times (-\tfrac{1}{2},-\tfrac{1}{2},-\tfrac{1}{2},-\tfrac{1}{2}): (0)\,.
  \end{aligned}  
  \label{k=2 RR}
\end{equation}
From \eqref{deg2} and \eqref{chis2}, we see that these states survive the orbifold projection for $n_1=0$ mod $6$, and have degeneracy 4. So, we find 4 vectors $(\pm1,0)$ and 4 scalars $(0)$ with the same mass and level-matching condition as in \eqref{hmk2} and \eqref{levelmatch2}. Similarly with the NS-NS (and NS-R) sector, the states in \eqref{k=2 RR} can also survive the orbifold projection for $n_1=3$ mod 6, and then, they have degeneracy 5. States in the R-NS sector are constructed similarly with the R-R sector. In the R-NS sector we find 8 dilatini $(\pm\tfrac{1}{2})$. In total, the lightest states from the R-R and R-NS sectors form 4 towers of vector multiplets with mass \eqref{hmk2}, where $n_1=0$ mod $6$, and 5 towers of vector multiplets with mass \eqref{hmk2}, where $n_1=3$ mod $6$. 

\subsubsection{$k=3$ sector}
Finally, we discuss the spectrum in the $k=3$ sector.  The shift in the zero point energies is (cf. \eqref{shifted energy}) 
\begin{equation}
    \widetilde{E}_3=E_3=\frac{1}{4}\,,
\end{equation}
and the degeneracy of a generic state is given by (cf. \eqref{fixed points}, \eqref{shift phase factor}, \eqref{degeneracy twisted})
\begin{equation}
    D(3)= \frac{1}{6}\sum_{l=0}^{5}{\chi} [3,l]\,\tilde{\chi}[3,l] e^{\frac{2\pi il}{6}[5\tilde{r}_3+\tilde{r}_4-(r_3+r_4) +n_1] }\,,
    \label{deg3}
\end{equation}
where
\begin{equation}
   {\chi} [3,l]\, \tilde{\chi}[3,l]= 
\begin{dcases}
   \: 16& \text{for}\quad l=0,3\,.\\
   \: 1& \text{for} \quad l=1,2,4,5\,.
\end{dcases}
\label{chis3}
\end{equation}
We list the weight vectors for the lightest left and right-moving states of the $k=3$ sector in \autoref{k=3 stu states}.
\renewcommand{\arraystretch}{2}
\begin{table}[h!]
\centering
 \begin{tabular}{|c|c|c|}
    \hline
    Sector &  $\tilde{r}$ &   ${r}$ \\
    \hline
    \hline
  \multirow{2}{*}{NS} & $(0,0,-3,0)$ &$(0,0,-1,0)$ \\
  \cline{2-3}
  & $(0,0,-2,-1)$ &$(0,0,0,-1)$ \\
  \hline
   \multirow{2}{*}{R}  & $(\frac{1}{2},\frac{1}{2},-\frac{5}{2},-\frac{1}{2})$& $(\frac{1}{2},\frac{1}{2},-\frac{1}{2},-\frac{1}{2})$ \\
   \cline{2-3}
   & $(-\frac{1}{2},-\frac{1}{2},-\frac{5}{2},-\frac{1}{2})$& $(-\frac{1}{2},-\frac{1}{2},-\frac{1}{2},-\frac{1}{2})$ \\
   \hline
    \end{tabular}
\captionsetup{width=.9\linewidth}
\caption{The weight vectors of the lightest left and right-moving states in the $k=3$ twisted sector.}
\label{k=3 stu states}
\end{table}
\renewcommand{\arraystretch}{1}

Let us start with the construction of states in the NS-NS sector, in which we find
\begin{equation}
    \begin{aligned}
       & (0,0,-3,0)\times (0,0,\underline{0,-1}): 2\times (0)\,,\\
       & (0,0,-2,-1)\times (0,0,\underline{0,-1}): 2\times (0)\,.
    \end{aligned}
    \label{k=3 NS-NS}
\end{equation}
As we can see from \eqref{deg3} and \eqref{chis3}, these states survive the orbifold projection for $n_1=0$ mod $6$, and have degeneracy 5. So, we find 20 scalars (0) with mass (cf. \eqref{twisted masses} with $\widetilde{N}=N=0$)
\begin{equation}
\begin{aligned}
    &\frac{\alpha'm^2_{\text{L}}(3)}{2}=\frac{1}{2}p^2_{\text{L}}(3)=\frac{1}{4T_2U_2}|n_2-Un_1 +\widebar{T}(w_1+\tfrac{1}{2}) +\widebar{T}Uw_2|^2\,,\\
    &\frac{\alpha'm^2_{\text{R}}(3)}{2}=\frac{1}{2}p^2_{\text{R}}(3)=\frac{1}{4T_2U_2}|n_2-Un_1 +{T}(w_1+\tfrac{1}{2}) +{T}Uw_2|^2\,,
\end{aligned}
    \label{hmk3}
\end{equation}
where $n_1=0$ mod $6$, and the level-matching condition reads
\begin{equation}
     n_1\left(w_1+\frac{1}{2}\right) + n_2w_2 = 0\,.
     \label{levelmatch3}
\end{equation}
As in the $k=2$ twisted sector, the states in \eqref{k=3 NS-NS} can also survive the orbifold projection for $n_1=2$ mod $6$ and $n_1=4$ mod $6$ and, in these cases, they come with degeneracy 5 or 6.

The construction of the lightest states in the NS-R sector is similar with the NS-NS sector, and we find 10 dilatini $(\pm\tfrac{1}{2})$ with mass \eqref{hmk3}. So, from the NS-NS and NS-R sectors we obtain 5 towers of hypermultiplets with mass \eqref{hmk3}, when $n_1=0$ mod $6$, and 11 towers of hypermultiplets with mass \eqref{hmk3}, from both $n_1=2$ mod $6$ and $n_1=4$ mod $6$.

Now, we move on to the R-R sector, in which we find
\begin{equation}
    \begin{aligned}
    &(\tfrac{1}{2},\tfrac{1}{2},-\tfrac{5}{2},-\tfrac{1}{2})\times (\tfrac{1}{2},\tfrac{1}{2},-\tfrac{1}{2},-\tfrac{1}{2}): (0)\,,\\
      &(\tfrac{1}{2},\tfrac{1}{2},-\tfrac{5}{2},-\tfrac{1}{2})\times (-\tfrac{1}{2},-\tfrac{1}{2},-\tfrac{1}{2},-\tfrac{1}{2}): (1)\,,\\
      &(-\tfrac{1}{2},-\tfrac{1}{2},-\tfrac{5}{2},-\tfrac{1}{2})\times (\tfrac{1}{2},\tfrac{1}{2},-\tfrac{1}{2},-\tfrac{1}{2}): (-1)\,,\\
      &(-\tfrac{1}{2},-\tfrac{1}{2},-\tfrac{5}{2},-\tfrac{1}{2})\times (-\tfrac{1}{2},-\tfrac{1}{2},-\tfrac{1}{2},-\tfrac{1}{2}): (0)\,.
    \end{aligned}
    \label{k=3 RR}
\end{equation}
From \eqref{deg3} and \eqref{chis3}, we can see that these states survive the orbifold projection for $n_1=0$ mod $6$, and have degeneracy 6. So, we find 6 vectors $(\pm1,0)$ and 6 scalars $(0)$ with the same mass and level-matching condition as in \eqref{hmk3} and \eqref{levelmatch3}. States in the R-NS sector are constructed similarly with the R-R sector. In the R-NS sector we find 12 dilatini $(\pm\tfrac{1}{2})$. In total, the lightest states from the R-R and R-NS sectors form 6 towers of vector multiplets with mass \eqref{hmk3} and $n_1=0$ mod 6. Finally, we mention here that the states in the R-R and R-NS sectors can also survive the orbifold projection with the addition of momentum number $n_1=2$ or $4$ mod $6$ and in these cases they come with degeneracy 5. So, we also find 10 towers of vector multiplets with mass \eqref{hmk3} and $n_1=2$ or $4$ mod $6$.

\section{Moduli space and the swampland}
\label{moduli and swampland}
In this section we first discuss the classical moduli space of the $STU$-like $\mathbb{Z}_6$ model studied in section \ref{z12 with nv=3 and nh=0}. Then we determine the locations in the moduli space where generically massive states become massless. There are points at infinite distance where infinite towers of states become massless, as well as special lines and points in the interior of the (classical) moduli space where only a finite number of states becomes massless.

\subsection{Classical moduli space}
In this subsection we will determine the vector multiplet and hypermultiplet moduli space of our $\mathbb{Z}_6$ orbifold model. First, we recall some classical aspects of the moduli spaces of the 5$D$ $\mathbb{Z}_6$ model studied in \cite{Gkountoumis:2024dwc}. In five dimensions, the vector multiplet moduli space is given by 
\begin{equation}\label{modspace2}
    {\cal M}_V^{(5)}=\mathbb{R}^+\times \mathbb{R}^+\ ,
\end{equation}
where $\mathbb{R}^+=\text{SO}(1,1)/\mathbb{Z}_2$. This moduli space is parametrized by the five-dimensional string coupling $\lambda_5$ and the radius of the circle $\mathcal{R}_5$ along which the fields have non-trivial monodromy. The corresponding $d$-symbols are (up to an overall normalization)
\begin{equation}\label{d-symbols1}
    d_{122}=1\ ,\qquad d_{133}=-1\ .
\end{equation}
The hypermultiplet scalars form the quaternion-K{\"a}hler manifold of real dimension 8 (with $\text{SU}(2,2)\simeq \text{SO}(4,2)$),
\begin{equation}
{\cal M}_H=\frac{\text{SU}(2,2)}{\text{SU}(2)\times \text{SU}(2)\times \text{U}(1)}\simeq \frac{\text{SO}(4,2)}{\text{SO}(4)\times \text{SO}(2)}\ .
\end{equation}
The moduli space of the $5D$ orbifold was found by computing the commutant of the twist matrix in the $5D$ T-duality group $\text{SO}(5,5)$. For this case, the commutant is \cite{Gkountoumis:2024dwc}
\begin{equation}
    \mathcal{C}^{(5)}=\text{SO}(1,1)\times \text{SU}(2,2)\times \text{U}(1)\,.
\end{equation}
In order to find the moduli space of the $4D$ model, we have to compute the commutant of the twist matrix in the $4D$ T-duality group $\text{SO}(6,6)$. But since the orbifold  acts as a symmetry of $T^4$ together with a shift on $S^1_{{\cal R}_5}$, it is straightforward to verify that the commutant is enhanced as follows
\begin{equation}
     \mathcal{C}^{(5)}\to \mathcal{C}^{(4)}=\text{SO}(2,2)\times \text{SU}(2,2)\times \text{U}(1)\ .
\end{equation}
Since the monodromy is in the T-duality group, it also commutes with the classical $4D$ S-duality group $\text{SL}(2)_S$. 
Then, using $\text{SO}(2,2)\sim \text{SL}(2) \times \text{SL}(2)$, the vector multiplet moduli space is
\begin{equation}\label{modspace3}
    {\cal M}_V^{(4)}= \frac{\text{SL}(2)}{\text{U}(1)}\times\frac{\text{SL}(2)}{\text{U}(1)}\times\frac{\text{SL}(2)}{\text{U}(1)}\,,
\end{equation}
and consists of 3 complex NS-NS scalar fields, which are the $S, \,T$ and $U$ moduli. This is all consistent with the supergravity dimensional reduction of the $5D$ theory to four dimensions, and is given by the so-called $r$-map. As we have explained in section \ref{duality groups of orbifolded theory}, the duality group of our $\mathbb{Z}_6$ orbifold is further broken to a subgroup
\begin{equation}
    \hat{\Gamma}^1(6)_S \times \hat{\Gamma}^1(6)_T \times \hat{\Gamma}_1(6)_U\,,
\end{equation}
due to the orbifold shift. Therefore, each of the $\text{SL}(2)$ factors in \eqref{modspace3} should be modded out by the corresponding congruence subgroup of $\mathrm{SL}(2,\mathbb{Z})$.

Now, the action of an $\mathcal{N}=2$ supergravity theory coupled to $n$ vector multiplets in $4D$ can be specified by the prepotential $F(X)$, which is a holomorphic and homogeneous function of second degree in the variables $X^I$, $I=0,\ldots,n$.
For the model at hand, the corresponding prepotential governing the $4D$ vector multiplets is given by 
\begin{equation}
    F(X)=id_{ABC}\frac{X^AX^BX^C}{X^0}\,.
\end{equation}
Here $A,B,C=1,2,3$, and the $d$-symbols are the same as in \eqref{d-symbols1}. So, we obtain
\begin{equation}
    F(X)= 3i\frac{X^1}{X^0}\left[(X^2)^2-(X^3)^2\right]\ .
\end{equation}
The complex coordinates $X^1,X^2,X^3$ are the complexifications of the $5D$ real coordinates denoted by $h^1,h^2,h^3$ in \cite{Gkountoumis:2024dwc}, as any $5D$ vector yields a scalar in $4D$ which pairs up with the $5D$ real scalar (reduced to $4D$) to make up the complex variables as part of the $4D$ vector multiplet \cite{Gunaydin:1983rk}. By applying a symplectic transformation we can bring the prepotential to the equivalent form 
\begin{equation}
     F(X)=i\frac{X^1X^2X^3}{X^0}\,.
\end{equation}
In this basis the only non-vanishing $d$-symbol is $d_{123}=1/6$. Identifying 
\begin{equation}
S=\frac{X^1}{X^0}\ ,\qquad  T=\frac{X^2}{X^0}\ ,\qquad U=\frac{X^3}{X^0}\ ,
\end{equation}
yields the classical prepotential
\begin{equation}
      F(X)= iSTU(X^0)^2\,.
      \label{classicalSTU}
\end{equation}
From the prepotential, one can also compute the K\"ahler potential $K$,  which is given by 
\begin{equation}
    K = -\ln Y = -\ln \left(\tfrac{1}{2}X^I\widebar{F}_I + \tfrac{1}{2}\widebar{X}^I{F}_I\right)\,,
    \label{defY}
\end{equation}
where $F_I=\partial F/\partial X^I$ and $I=0,\dots,3$.
 In our case, we find  
\begin{equation}
    Y = \frac{1}{2i}(S-\widebar{S})(T-\widebar{T})(U-\widebar{U})\left|X^0\right|^2\,.
    \label{kahlerY}
\end{equation}
The vector multiplet moduli space is a K\"ahler manifold, and its metric is determined by the second derivative of the K\"ahler potential \eqref{defY}, which is
\begin{equation}
\begin{aligned}
    K_{I\bar{J}}\ \rmd z^I \rmd\widebar{z}^{\bar{J}}&=-\frac{1}{(S-\widebar{S})^2}\rmd S\rmd \widebar{S} - \frac{1}{(T-\widebar{T})^2}\rmd T\rmd \widebar{T} - \frac{1}{(U-\widebar{U})^2}\rmd U\rmd \widebar{U}\\&=\frac{1}{4S_2^2}\left(\rmd S_1^2+\rmd S_2^2\right)+\frac{1}{4T_2^2}\left(\rmd T_1^2+\rmd T_2^2\right)+\frac{1}{4U_2^2}\left(\rmd U_1^2+\rmd U_2^2\right)\\
    &=\frac{1}{4}e^{-2\sqrt{2}\phi_S}\rmd S_1^2 +\frac{1}{2}\rmd\phi_S^2 + \frac{1}{4}e^{-2\sqrt{2}\phi_T}\rmd T_1^2 +\frac{1}{2}\rmd\phi_T^2+\frac{1}{4}e^{-2\sqrt{2}\phi_U}\rmd U_1^2 +\frac{1}{2}\rmd\phi_U^2.
\end{aligned}
\end{equation}
Here $\phi_I,\,I=S,T,U$, are the corresponding normalized moduli, defined by
\begin{equation}
    S_2 = e^{\sqrt{2}\phi_S},\quad T_2 = e^{\sqrt{2}\phi_T},\quad U_2 = e^{\sqrt{2}\phi_U}.
    \label{normalizedmoduli}
\end{equation}
These give the normalized classical moduli space metric.

\subsection{Massless states at finite distance}
\label{Massless states at finite distance}
In this section we will analyse the special lines and points at finite distance in the interior of the moduli space where massive states become massless. It is important to mention here that in the interior of the moduli space only a finite number of states become massless and no infinite towers become massless. 

Recall that in  the $k=1$ sector we found half a hypermultiplet\footnote{The other half arises in the $k=5$ sector.} with mass
\begin{equation}
\begin{aligned}
    &\frac{\alpha'm^2_{\text{L}}(1)}{2}=\frac{1}{2}p^2_{\text{L}}(1)-\frac{1}{3}=\frac{1}{4T_2U_2}|n_2-Un_1 +\widebar{T}(w_1+\tfrac{1}{6}) +\widebar{T}Uw_2|^2-\frac{1}{3}\,,\\
    &\frac{\alpha'm^2_{\text{R}}(1)}{2}=\frac{1}{2}p^2_{\text{R}}(1)=\frac{1}{4T_2U_2}|n_2-Un_1 +{T}(w_1+\tfrac{1}{6}) +{T}Uw_2|^2\,,
\end{aligned}
\label{hmk1new}
\end{equation}
with the constraint $n_1=2$ mod $6$, and the level-matching condition
\begin{equation}
    n_1\left(w_1+\frac{1}{6}\right) + n_2w_2 = \frac{1}{3}\,.
    \label{levelhmk1new}
\end{equation}
From the mass formulae \eqref{hmk1new}, we see that massless states can appear if
\begin{equation}
    T =\frac{n_1U-n_2}{w_2U+(w_1+\tfrac{1}{6})}\,,
\end{equation}
where $n_1=2$ mod $6$ and the level-matching constraint \eqref{levelhmk1new} should be satisfied. We can rewrite the above expression together with the level-matching condition in a more convenient form as follows
\begin{equation}
    \frac{T}{6} =\frac{n_1U-n_2}{6w_2U+(6w_1+1)}\,,\qquad \frac{n_1}{2}(6w_1+1) + 3n_2w_2 = 1\,.
\end{equation}
By setting $\alpha=\frac{n_1}{2}\in 1+3\mathbb{Z}$,  $\beta=-n_2\in \mathbb{Z}$, $\gamma=3w_2\in 3\mathbb{Z}$ and $\delta=1+6w_1\in 1+6\mathbb{Z}$, we obtain
\begin{equation}
    \frac{T}{6}=\frac{\alpha2U+\beta}{\gamma 2U+\delta}\,,\qquad \alpha\delta-\beta\gamma=1\,.
\end{equation}
Then massless states will appear when
\begin{equation}
    \frac{T}{6} = g(2U)\,,\qquad  g\in  \mathcal{S}\equiv \left\{\alpha\delta-\beta\gamma=1:\,\alpha\in 1+3\mathbb{Z},\,\beta\in\mathbb{Z},\,\gamma\in3\mathbb{Z},\,\delta\in1+6\mathbb{Z}\,\right\}.
     \label{t12branch}
\end{equation}
It can be easily proven that all solutions of \eqref{t12branch} can be obtained from the solution  $\frac{T}{6}=2U$ by applying $\Gamma^1(6)_T\times \Gamma_1(6)_U$ transformations. Starting from the solution $\frac{T}{6}=2U$ and applying a generic $\Gamma^1(6)_T\times \Gamma_1(6)_U$ transformation we obtain (cf. \eqref{gamma1t} and \eqref{gamma1u})
\begin{equation}
    \frac{a\frac{T}{6}+\frac{b}{6}}{cT+d}=\frac{a'2U+2b'}{c'U+d'}\implies \frac{a\frac{T}{6}+\frac{b}{6}}{6c\frac{T}{6}+d}=\frac{a'2U+2b'}{\frac{c'}{2}2U+d'}\,.
\end{equation}
Here $a,d,a',d'\in 1+6\mathbb{Z}$, $b,c'\in6\mathbb{Z}$ and $c,b'\in\mathbb{Z}$. Solving for $\frac{T}{6}$ yields
\begin{equation}
    \frac{T}{6} = \frac{(da'-\frac{bc'}{12})2U+(2db'-\frac{bd'}{6})}{(\frac{ac'}{2}-6a'c)2U+(ad'-12b'c)}\,.
    \label{t12ug1tg1u}
\end{equation}
Now, we note that
\begin{equation}
\begin{aligned}
    &\alpha\equiv da'-\frac{bc'}{12}\in1+3\mathbb{Z}\,,\qquad \beta \equiv 2db'-\frac{bd'}{6} \in \mathbb{Z}\,,\\
    &\gamma \equiv \frac{ac'}{2}-6a'c \,\in \,3\mathbb{Z}\,,\qquad \delta \equiv ad'-12b'c \,\in 1+6\mathbb{Z}\,,\qquad\alpha \delta - \beta\gamma = 1\,.
    \end{aligned}
\end{equation}
So, we can rewrite \eqref{t12ug1tg1u} as
\begin{equation}
     \frac{T}{6}=\frac{\alpha2U+\beta}{\gamma 2U+\delta}\,,\qquad \alpha\delta-\beta\gamma=1\,,
\end{equation}
which exactly  reproduces \eqref{t12branch}. 

As we have already discussed, the sectors $k=1$ and $k=5$ are equivalent. So, we conclude that at the critical line $\frac{T}{6}=2U$, modulo $\Gamma^1(6)_T\times \Gamma_1(6)_U$ transformations, 1 hypermultiplet becomes massless.

\begin{figure}[h]
\centering
\includegraphics[width=0.5\textwidth]{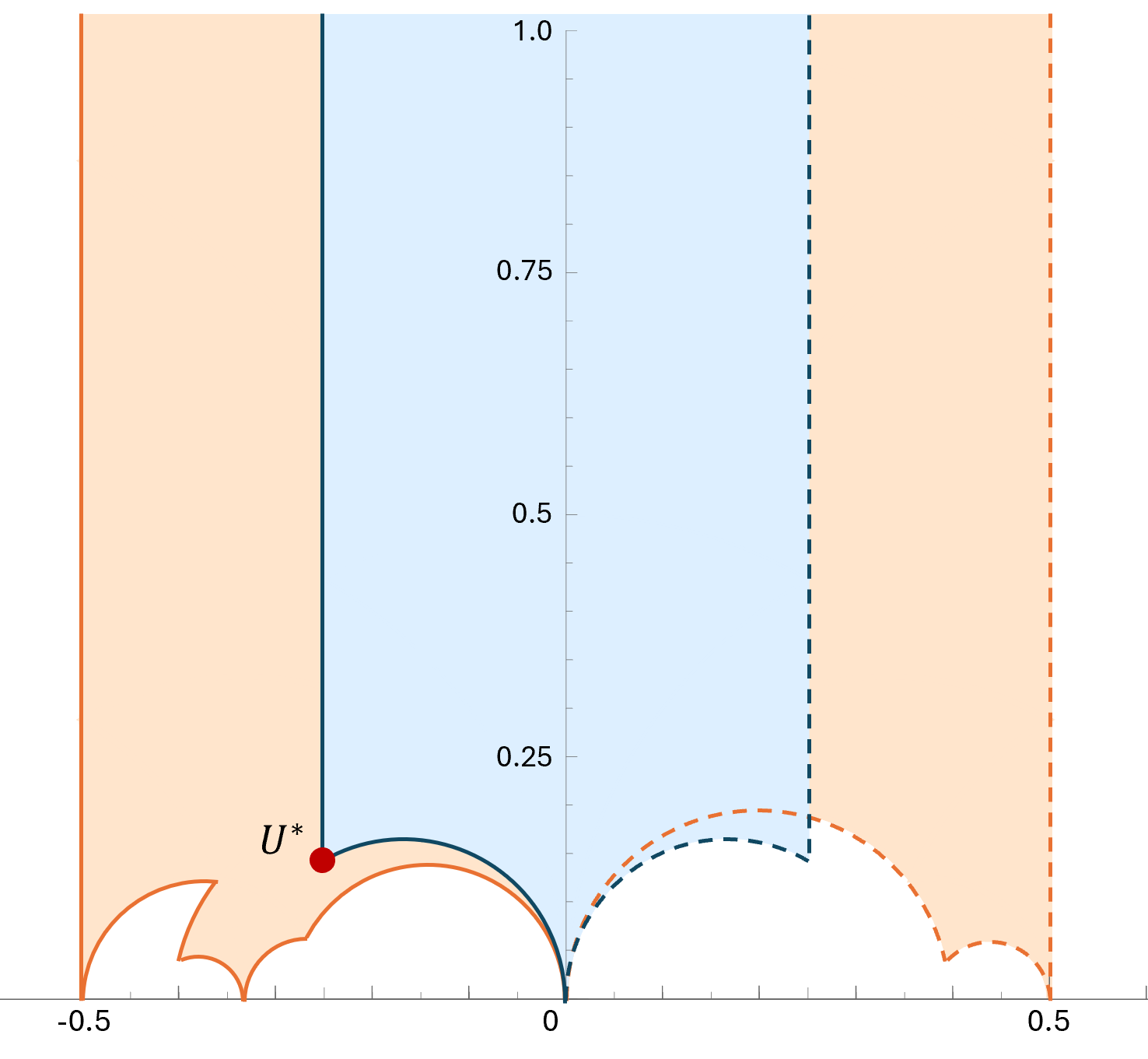}
\caption{The location of $U^*$, represented by the red dot. The orange region is the fundamental domain of $\Gamma_1(6)_U$, whereas the blue region is the fundamental domain of $\Gamma_1(3)_{2U}$. The shapes of the fundamental domains are obtained via \texttt{DrawFunDoms.m}.}
\label{funddom63}
\end{figure}

Now, as in  heterotic string constructions (see e.g. \cite{LopesCardoso:1994ik}), whenever two (or more) critical lines intersect, two (or more) hypermultiplets will become massless. Starting from \eqref{t12branch}, it is easy to show that the only lines intersecting inside the fundamental domain of $\Gamma^1(6)_T\times \Gamma_1(6)_U$ are the lines $ \frac{T}{6}=2U$ and $\frac{T}{6}=-(4U+1)/(6U+1)$. These lines intersect at the point 
$(T,U)=(12U^*,U^*)$, where $U^*= -\frac{1}{4}+i\frac{\sqrt{3}}{12}$, and it is depicted in \autoref{funddom63}. Hence, at this point two hypermultiplets become massless.

Finally, in the $k=1$ sector we also found one tower of hypermultiplets with mass
\begin{equation}
\begin{aligned}
    &\frac{\alpha'm^2_{\text{L}}(1)}{2}=\frac{1}{2}p^2_{\text{L}}(1)-\frac{1}{3}+\frac{1}{6}=\frac{1}{4T_2U_2}|n_2-Un_1 +\widebar{T}(w_1+\tfrac{1}{6}) +\widebar{T}Uw_2|^2-\frac{1}{6}\,,\\
    &\frac{\alpha'm^2_{\text{R}}(1)}{2}=\frac{1}{2}p^2_{\text{R}}(1)=\frac{1}{4T_2U_2}|n_2-Un_1 +{T}(w_1+\tfrac{1}{6}) +{T}Uw_2|^2\,,
\end{aligned}
\label{hmk1-case2new}
\end{equation}
with the constraint $n_1=1$ mod $6$ and the level-matching condition 
\begin{equation}
    n_1\left(w_1+\frac{1}{6}\right) + n_2w_2 = \frac{1}{6}\,.
    \label{levelhmk1-case2new}
\end{equation}
From the mass formulae \eqref{hmk1-case2new}, we see that massless states can appear if
\begin{equation}
    T =\frac{n_1U-n_2}{w_2U+(w_1+\tfrac{1}{6})}\,,
\end{equation}
where $n_1=1$ mod $6$ and the level-matching constraint \eqref{levelhmk1-case2new} should be satisfied. Again, we can rewrite the above expression together with the level-matching condition in a more convenient form as follows
\begin{equation}
    \frac{T}{6} =\frac{n_1U-n_2}{6w_2U+(6w_1+1)}\,,\qquad n_1(6w_1+1) + 6n_2w_2 = 1\,.
\end{equation}
By setting $a=n_1\in 1+6\mathbb{Z}$, $b=-n_2\in \mathbb{Z}$, $c=6w_2\in 6\mathbb{Z}$, and $d=1+6w_1\in 1+6\mathbb{Z}$ we obtain
\begin{equation}
   \frac{T}{6} =\frac{a{U}+b}{c{U}+d}\,,\qquad ad-bc=1\,.
   \label{Tmod6U}
\end{equation}
So, massless states will appear when
\begin{equation}
   \frac{T}{6}= gU\,,\qquad g\in \Gamma_1(6)_U\,.
\end{equation}
 or, equivalently,
 \begin{equation}
      {hT}= 6U\,,\qquad h\in \Gamma^1(6)_T\,.
 \end{equation}
We mention here that for the branch of solutions \eqref{Tmod6U}, there are no intersecting lines. As we have already mentioned the sectors $k=1$ and $k=5$ are equivalent. Thus, at $\frac{T}{6}=U$ (mod $\Gamma^1(6)_T\times \Gamma_1(6)_U$) we get another massless hypermultiplet from the $k=5$ sector. 

Concluding, we have found that there exist 2 special lines and 1 special point in the bulk of the moduli space, where massive hypermultiplets become massless. In particular, at $\frac{T}{6}=U$, modulo $\Gamma^1(6)_T\times \Gamma_1(6)_U$ transformations, 2 hypermultiplets become massless, and at $\frac{T}{6}=2U$, modulo $\Gamma^1(6)_T\times \Gamma_1(6)_U$ transformations, 1 hypermultiplet becomes massless. At the special point $(T,U)=(12U^*,U^*)$, where $U^*= -\frac{1}{4}+i\frac{\sqrt{3}}{12}$, two critical lines intersect, and 2 hypermultiplets become massless. Note that  the hypermultiplets that become massless carry momentum and winding numbers along the $T^2$ directions, so they are charged under the corresponding vector fields.

The appearance of massless states in the bulk of the moduli space has significant consequences for the structure of the moduli space at the quantum level. As it was shown in \cite{deWit:1995dmj, Antoniadis:1995ct}, the one-loop prepotential exhibits logarithmic singularities exactly at the lines in the moduli space where generically massive, charged states become massless. Hence, the classical moduli space is modified by quantum corrections. The computation of such corrections in our model relies on modularity properties of the prepotential under the congruence subgroups of $\text{SL}(2;\mathbb{Z})$. We will discuss this issue in detail in an upcoming work.

\subsection{Massless states at infinite distance}\label{Massless states at infinite distance}
In this section we will study the spectrum of states at all different asymptotic limits in the $T-U$ plane of the moduli space. First, we will discuss these limits separately for $T$ and $U$, i.e. we will fix $U$ in the bulk of the moduli space and we will consider the infinite distance points in the $T$-plane and vice-versa. These will be refered to as single cusps. Then we will consider the asymptotic limits simultaneously for both $T$ and $U$, which will be refered to as double cusps.

\subsubsection{Single cusps as 6$D$ limits}

\begin{figure}[h]
\centering
\begin{subfigure}{.7\textwidth}
  \centering
  \includegraphics[width=.86\linewidth]{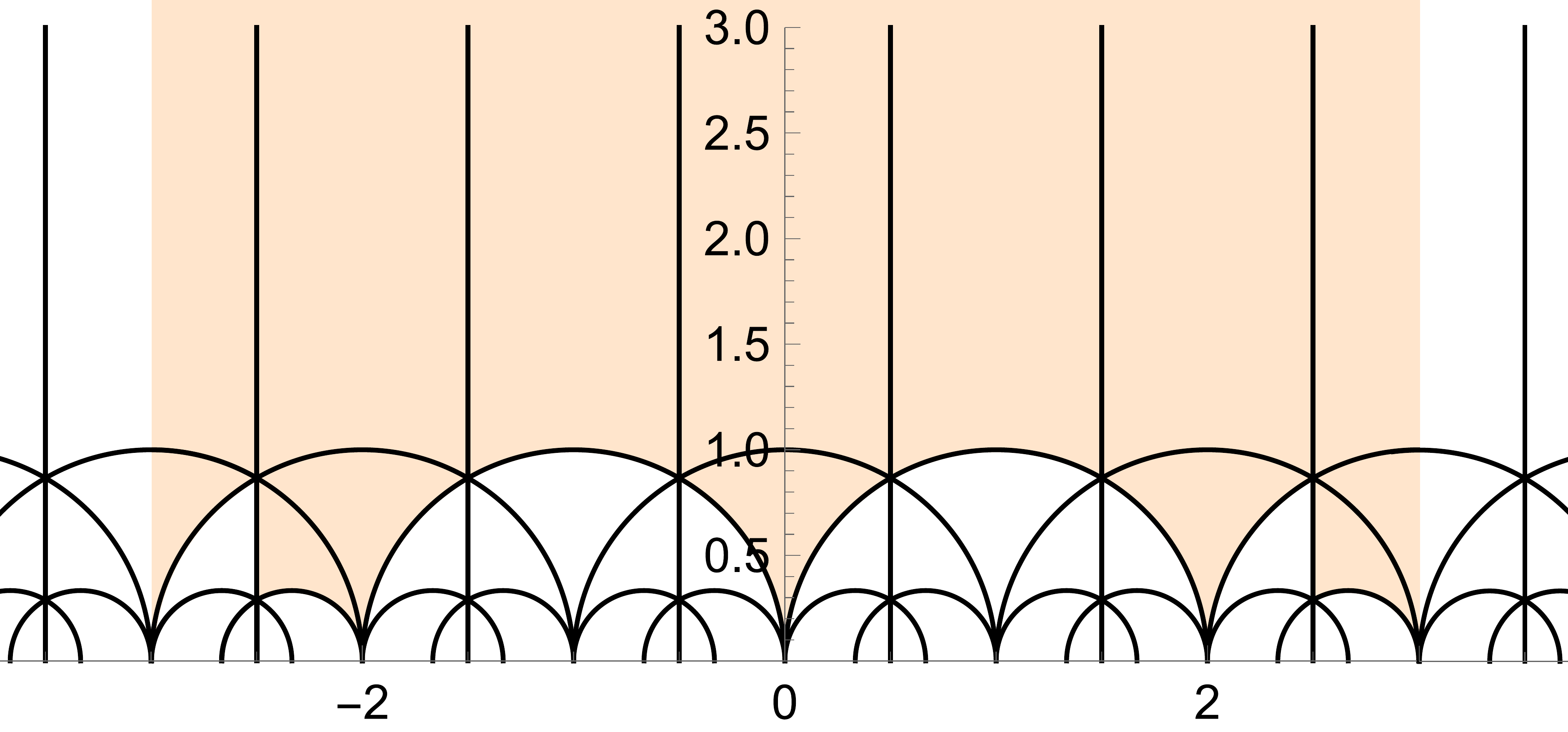}
\end{subfigure}%
\begin{subfigure}{.3\textwidth}
  \centering
  \includegraphics[width=0.75\linewidth]{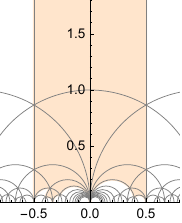}
\end{subfigure}
\caption{The fundamental domains of the congruence subgroups $\Gamma^1(6)_T$ and $\Gamma_1(6)_U$. The shaded region of the left and right figure corresponds to the fundamental domain of $\Gamma^1(6)_T$ and $\Gamma_1(6)_U$, respectively. The figures are obtained by the Mathematica package \texttt{DrawFunDoms.m}.}
\label{funddom}
\end{figure}

The fundamental domains of $\Gamma^1(6)_T$ and $\Gamma_1(6)_U$ are shown in \autoref{funddom}. There are 4 inequivalent cusps in the fundamental domain of $\Gamma^1(6)_T$, at the points $T=-3,-2,0$ and $i\infty$. Similarly, there exist 4 inequivalent cusps in the fundamental domain of $\Gamma_1(6)_U$, at the points $U=-\frac{1}{2},-\frac{1}{3},0$ and $i\infty$.

\renewcommand{\arraystretch}{2}
\renewcommand{\arraystretch}{1}

The cusp points correspond to singular geometric structures of the torus. In the limit $T_2\to 0 $, with $U$ constant and finite, the volume of $T^2$ goes to zero, while the complex structure remains regular. This implies that the torus shrinks to a point. For the limit $U_2\to 0 $, with $T$ constant and finite, the torus, as an elliptic curve, has zero discriminant and divergent modular invariants, so that it becomes a nodal curve. In the special case in which $U_1=0$, the nodal curve is described by  $\mathcal{R}_4\to0$ and  $\mathcal{R}_5\to\infty$. 

We should highlight here that all cusp points are at the asymptotic boundary of the fundamental domain. Hence, it is worth studying the spectrum and physical interpretation around these infinite distance points, and investigating whether the Swampland Distance Conjecture \cite{Ooguri:2006in} is valid. The Swampland Distance Conjecture (SDC) proposes that as we approach an asymptotic limit on the moduli space, there must be at least one tower of states, whose mass scale decreases exponentially as
\begin{equation}\label{SDC}
    m= m_0 e^{-\lambda |\Delta\phi|}\,,
\end{equation}
where $\phi$ is a normalized modulus and $\lambda\sim\mathcal{O}(1)$. To verify this conjecture, let us analyse the spectrum of lightest states of all sectors with $N=\widetilde{N}=0$.

The mass formulae for the lightest states that we found in the previous section  (cf. \eqref{toweruntwisted}, \eqref{vmk1}, \eqref{hmk2}, and \eqref{hmk3}) can be summarized as follows:
\begin{equation}
\begin{aligned}
    &\frac{\alpha'm^2_{\text{L}}(k)}{2}=\frac{1}{4T_2U_2}\left|n_2-Un_1 +\widebar{T}\left(\hat{w}_1 +Uw_2\right)\right|^2=\frac{1}{4T_2U_2}\left|n_2+T\hat{w}_1+\widebar{U}\left(-n_1+Tw_2\right)\right|^2\,,\\
    &\frac{\alpha'm^2_{\text{R}}(k)}{2}=\frac{1}{4T_2U_2}\left|n_2-Un_1 +{T}\left(\hat{w}_1 +Uw_2\right)\right|^2=\frac{1}{4T_2U_2}\left|n_2+T\hat{w}_1+U\left(-n_1+Tw_2\right)\right|^2\,,
\end{aligned}
    \label{mk}
\end{equation}
and the level matching condition reads
\begin{equation}
    n_1 \hat{w}_1+n_2 w_2=0\,,\quad \text{where }\quad \hat{w}_1\equiv w_1 +\frac{k}{6}\,, \qquad k=0,1,\ldots,5\,.
\end{equation}
Utilising this level matching condition, it can be derived from \eqref{mk} that
\begin{equation}\label{mkfinal}
\alpha'm^2(k)=\frac{1}{T_2U_2w_2^2}\left|-n_1+Tw_2\right|^2\left|\hat{w}_1+Uw_2\right|^2=\frac{1}{T_2U_2n_2^2}\left|n_2+T\hat{w}_1\right|^2\left|-n_2+Un_1\right|^2\,.
\end{equation}
Given these, we can solve the equation $m^2=m^2_{\text{L}}(k)+m^2_{\text{R}}(k)=0$, as we approach the infinite distance points in the $T-U$ moduli space. There are 4 inequivalent sets of solutions: 
\begin{itemize}
    \item $T_2\to0$: $T=T_1=\frac{n_1}{w_2}=-\frac{n_2}{\hat{w}_1}$, while $U$ is in the bulk.
    \item $U_2\to0$: $U=U_1=\frac{n_2}{n_1}=-\frac{\hat{w}_1}{w_2}$, while $T$ is in the bulk.
    \item $T_2\to\infty$: $\hat{w}_1=w_2=0$, while $U$ is in the bulk.
    \item $U_2\to\infty$: $n_1=w_2=0$, while $T$ is in the bulk.
\end{itemize}

\renewcommand{\arraystretch}{1.7}
\begin{table}[t!]
\centering
 \begin{tabular}[c]{|c|l|l|c|c|}
    \hline
    Sector & \multicolumn{2}{|c|}{Cusp} & Lattice constraints & Mass of tower \\
    \hline
    \hline
  \multirow{6}{*}{$k=0$}& \multirow{3}{*}{$T_2\to0$} & $T_1=0$ &$n_1=n_2=0$ & \multirow{3}{*}{$\frac{1}{\sqrt{\alpha'U_2}}\left|w_1+Uw_2\right|\sqrt{T_2}$} \\
  \cline{3-4}
  &  & $T_1=-2$ & $n_1=-2w_2,\,n_2=2w_1$ &  \\
  \cline{3-4}
  & & $T_1=-3$ & $n_1=-3w_2,\,n_2=3w_1$ & \\
  \cline{2-5}
  & \multirow{3}{*}{$U_2\to0$} & $U_1=0$ & $n_2=w_1=0$ & \multirow{3}{*}{$\frac{1}{\sqrt{\alpha'T_2}}\left|n_1-Tw_2\right|\sqrt{U_2}$}\\
  \cline{3-4}
  & & $U_1=-\frac{1}{3}$ & $n_1=-3n_2,\,w_2=3w_1$ & \\
  \cline{3-4}
  & & $U_1=-\frac{1}{2}$ & $n_1=-2n_2,\,w_2=2w_1$ & \\
  \hline
   \multirow{1}{*}{$k=1$} & $T_2\to0$ & $T_1=0$& $n_1=n_2=0$ & $\frac{1}{\sqrt{\alpha'U_2}}\left|w_1+\frac{1}{6}+Uw_2\right|\sqrt{T_2}$ \\
   \hline
   \multirow{3}{*}{$k=2$} & \multirow{2}{*}{$T_2\to0$} & $T_1=0$& $n_1=n_2=0$ & \multirow{2}{*}{$\frac{1}{\sqrt{\alpha'U_2}}\left|w_1+\frac{1}{3}+Uw_2\right|\sqrt{T_2}$} \\
   \cline{3-4}
   & & $T_1=-3$& $n_1=-3w_2,\,n_2=3w_1+1$ &  \\
   \cline{2-5}
   &$U_2\to0$ & $U_1=-\frac{1}{3}$& $n_1=-3n_2,\,w_2=3w_1+1$ & $\frac{1}{\sqrt{\alpha'T_2}}\left|n_1-Tw_2\right|\sqrt{U_2}$ \\
   \hline
   \multirow{3}{*}{$k=3$} & \multirow{2}{*}{$T_2\to0$} & $T_1=0$& $n_1=n_2=0$ & \multirow{2}{*}{$\frac{1}{\sqrt{\alpha'U_2}}\left|w_1+\frac{1}{2}+Uw_2\right|\sqrt{T_2}$} \\
   \cline{3-4}
   & & $T_1=-2$& $n_1=-2w_2,\,n_2=2w_1+1$ & \\
   \cline{2-5}
   &$U_2\to0$ & $U_1=-\frac{1}{2}$& $n_1=-2n_2,\,w_2=2w_1+1$ & $\frac{1}{\sqrt{\alpha'T_2}}\left|n_1-Tw_2\right|\sqrt{U_2}$ \\
   \hline
   \multirow{3}{*}{$k=4$} & \multirow{2}{*}{$T_2\to0$} & $T_1=0$& $n_1=n_2=0$ & \multirow{2}{*}{$\frac{1}{\sqrt{\alpha'U_2}}\left|w_1+\frac{2}{3}+Uw_2\right|\sqrt{T_2}$} \\
   \cline{3-4}
   & & $T_1=-3$& $n_1=-3w_2,\,n_2=3w_1+2$ &  \\
   \cline{2-5}
   &$U_2\to0$ & $U_1=-\frac{1}{3}$& $n_1=-3n_2,\,w_2=3w_1+2$ & $\frac{1}{\sqrt{\alpha'T_2}}\left|n_1-Tw_2\right|\sqrt{U_2}$ \\
   \hline
   \multirow{1}{*}{$k=5$} & $T_2\to0$ & $T_1=0$& $n_1=n_2=0$ & $\frac{1}{\sqrt{\alpha'U_2}}\left|w_1+\frac{5}{6}+Uw_2\right|\sqrt{T_2}$ \\
   \hline
    \end{tabular}
\captionsetup{width=.9\linewidth}
\caption{All towers of states that become massless for $T_2\to0$ and $U$ fixed in the bulk of the moduli space, or $U_2\to 0$ and $T$ fixed in the bulk of the moduli space.}
\label{cusptowers}
\end{table}
\renewcommand{\arraystretch}{1}

In \autoref{cusptowers}, we list all towers of states that become massless as $T_2\to0$ or $U_2\to 0$. Note that there are multiple towers of states becoming massless as $T_2\to 0$ or $U_2\to0$. By using \eqref{normalizedmoduli}, comparing \eqref{SDC} with the last column of \autoref{cusptowers} and identifying $|\Delta\phi|=|\phi_T|$ or $|\phi_U|$, we can see that for all towers $\lambda = \frac{1}{\sqrt{2}}$. Then, as $T_2\to0$ or $U_2\to 0$, that is $\phi_T\to -\infty$ or $\phi_U\to -\infty$ the masses of all towers decrease as
\begin{equation}
     m= m_0 e^{-\frac{1}{\sqrt{2}} |\phi_I|}\,,\qquad I=T,U\,.
\end{equation}
In the limit $T_2\to \infty$, that is $\phi_T\to \infty$, it is easy to see that there exist towers of KK states only in the untwisted sector whose masses decay as
\begin{equation}
  m=  \frac{1}{\sqrt{\alpha'U_2}}\left|n_2-Un_1\right| e^{-\frac{1}{\sqrt{2}} |\phi_T|}\,.
\end{equation}
Finally, in the limit $U_2\to \infty$, namely $\phi_U\to\infty$, there exist towers of states in all sectors whose masses decay exponentially fast. For example, in the untwisted sector we find towers with mass
\begin{equation}
     m=  \frac{1}{\sqrt{\alpha'T_2}}\left|n_2+Tw_1\right| e^{-\frac{1}{\sqrt{2}} |\phi_U|}\,.
\end{equation}
Hence, regarding the single cusps on the $T-U$ moduli space, the SDC is satisfied  due to the existence of all the aforementioned towers of states.

The Emergent String Conjecture \cite{Lee:2019wij} proposes that each infinite distance limit on the moduli space 
corresponds to either a decompactification in which an infinite tower of Kaluza-Klein modes become massless, or a limit in which a string becomes tensionless and an infinite tower of string modes become massless. For our compactification, we will demonstrate that the infinite distance limits on the $T-U$ moduli space are all decompactification limits.

\subsubsection*{Asymptotic supersymmetry enhancement}
In our 4-dimensional $\mathcal{N}=2$ theory, there are 6 massive gravitini and 2 massless gravitini. These gravitini are in the multiplets of the untwisted sector. The massless gravitini arise in the NS-R sector and survive the orbifold projection if $n_1=0$ mod $6$. The massive gravitini come from the NS-R and R-NS sector and carry non-trivial orbifold charge, as can be easily checked by using \autoref{untwisted stu states} and the formula \eqref{degeneracy untwisted}. This charge can be cancelled by the addition of an appropriate momentum number $n_1$, which makes the gravitini massive. This indicates that supersymmetry is spontaneously broken from $\mathcal{N}=8$ to $\mathcal{N}=2$. The weight vectors of the 6 massive gravitini and their corresponding momentum numbers are:
\begin{equation}\label{massivegravitini}
    \begin{aligned}
    & (\pm 1,0,0,0)\times (\pm\tfrac{1}{2},\pm\tfrac{1}{2},\tfrac{1}{2},\tfrac{1}{2}) :  \left(\pm\tfrac{3}{2},\pm\tfrac{1}{2}\right)\,,\quad & n_1=1\bmod 6\,,\\
    & (\pm 1,0,0,0)\times (\pm\tfrac{1}{2},\pm\tfrac{1}{2},-\tfrac{1}{2},-\tfrac{1}{2}) :  \left(\pm\tfrac{3}{2},\pm\tfrac{1}{2}\right)\,,\quad & n_1=5\bmod 6\,,\\
    & (\pm\tfrac{1}{2},\pm\tfrac{1}{2},\tfrac{1}{2},\tfrac{1}{2})\times (\pm 1,0,0,0) : \left(\pm\tfrac{3}{2},\pm\tfrac{1}{2}\right)\,,\quad & n_1=3\bmod 6\,,\\
    & (\pm\tfrac{1}{2},\pm\tfrac{1}{2},-\tfrac{1}{2},-\tfrac{1}{2})\times (\pm 1,0,0,0) :  \left(\pm\tfrac{3}{2},\pm\tfrac{1}{2}\right)\,,\quad & n_1=3\bmod 6\,,\\
       & (\pm\tfrac{1}{2},\mp\tfrac{1}{2},\tfrac{1}{2},-\tfrac{1}{2})\times (\pm 1,0,0,0) : \left(\pm\tfrac{3}{2},\pm\tfrac{1}{2}\right)\,,\quad & n_1=4\bmod 6\,,\\
    & (\pm\tfrac{1}{2},\mp\tfrac{1}{2},-\tfrac{1}{2},\tfrac{1}{2})\times (\pm 1,0,0,0) : \left(\pm\tfrac{3}{2},\pm\tfrac{1}{2}\right)\,,\quad & n_1=2\bmod 6\,. 
    \end{aligned}
\end{equation}
In summary, 1 gravitino survives the orbifold projection if $n_1=1$ mod $6$, 1 gravitino survives if $n_1=2$ mod $6$, 2 gravitini survive if $n_1=3$ mod $6$,  1 gravitino survives if $n_1=4$ mod $6$ and  1 gravitino survives if $n_1=5$ mod $6$.

The masses of the gravitini can be easily computed using \eqref{untwisted masses}. We find 
\begin{equation}
\begin{aligned}
    &\frac{\alpha'm^2_{\text{L}}(0)}{2}=\frac{1}{2}p^2_{\text{L}}(0)=\frac{1}{4T_2U_2}|n_2-Un_1 +\widebar{T}w_1 +\widebar{T}Uw_2|^2\,,\\
    &\frac{\alpha'm^2_{\text{R}}(0)}{2}=\frac{1}{2}p^2_{\text{R}}(0)=\frac{1}{4T_2U_2}|n_2-Un_1 +{T}w_1 +{T}Uw_2|^2\,.
    \label{gravitinimasses}
\end{aligned}
 \end{equation}
Also, the level-matching condition reads
\begin{equation}
    n_1w_1+n_2w_2=0\,.
    \label{levelgravitini}
\end{equation}
It is important to stress here that the quantum number $n_1$ in \eqref{gravitinimasses} and \eqref{levelgravitini} is constrained, according to \eqref{massivegravitini}.

Now, as we approach the various cusps in the $T-U$ moduli space, all gravitini may remain massive, or some (or all) of them may become massless, as can be easily verified by using \eqref{massivegravitini}, \eqref{gravitinimasses} and \autoref{cusptowers}. In addition, all these infinite distance points can be interpreted as decompactification limits in type IIB theory or in a dual type IIA picture. These limits can be understood by studying the behaviour of the $T^2$ partition function at all cusps; a detailed analysis can be found in \autoref{details on partition app}. This is consistent with the Gravitini Mass (Swampland) Conjecture \cite{Cribiori:2021gbf,Castellano:2021yye}. This conjecture proposes that a massive gravitino can only become massless with an infinite tower of states at an infinite distance on the moduli space, and the gravitino mass is proportional to some power of the KK (or string excitation) mass scale. The power is between 1 and 3; in our case, it is 1.

Moreover, the effective supergravity theory becomes six-dimensional. The massless spectrum follows immediately from the analysis of section \ref{z12 with nv=3 and nh=0}, with the only difference that all fields fall in representations of the massless little group $\text{SU}(2)\times \text{SU}(2)$ in $6D$. For the supergravity multiplets in $6D$ and our conventions we refer to \autoref{6dsugra}.

We first discuss the cusps of the modulus $T$ and we keep $U$ fixed:
\begin{itemize}
    \item $T\to i\infty$: All gravitini become massless and supersymmetry in enhanced from $\mathcal{N}=2$ to $\mathcal{N}=8$. The resulting theory is  type IIB theory on $\mathbb{R}^{1,5}\times T^4$. The massless fields make up the $\mathcal{N}=8$ gravity multiplet is $6D$.
    \item $T\to 0$: All gravitini remain massive. The theory becomes type IIB on a non-freely acting asymmetric orbifold $\mathbb{R}^{1,5}\times T^4/\mathbb{Z}_6$, characterized by the twist vectors  $\tilde{u}=\left(\frac{5}{6},\frac{1}{6}\right)$ and $u=\left(\frac{1}{6},\frac{1}{6}\right)$. The massless fields make up the $\mathcal{N}=2$ gravity multiplet in $6D$ coupled to $9$ tensor multiplets, $8$ vector multiplets and $20$ hypermultiplets, satisfying the gravitational anomaly cancellation condition $n_H-n_V=273-29n_T$.
    \item $T\to -2$: The two R-NS gravitini carrying momentum number $n_1 = 2\bmod 6$ and $n_1 = 4\bmod 6$ become massless and supersymmetry in enhanced from $\mathcal{N}=2$ to $\mathcal{N}=4$. At this cusp we obtain type IIB on a non-freely acting symmetric orbifold $\mathbb{R}^{1,5}\times T^4/\mathbb{Z}_2$, characterized by the twist vectors $\tilde{u}=u=\left(\frac{1}{2},\frac{1}{2}\right)$. The massless fields make up the $\mathcal{N}=4\,(0,2)$ gravity multiplet and 21 tensor multiplets in $6D$. The same spectrum could also be obtained from type IIB on $\mathbb{R}^{1,5}\times \mathrm{K}3$. Also, note that the resulting number of tensor multiplets is exactly the number that is required for the gravitational anomalies to cancel in a chiral $\mathcal{N}=4\,(0,2)$ theory in 6$D$ \cite{townsend1984new}. 
    \item $T\to -3$: The two R-NS gravitini carrying momentum number $n_1 = 3 \bmod 6$ become massless and supersymmetry in enhanced from $\mathcal{N}=2$ to $\mathcal{N}=4$. In this case, we get type IIB theory on a non-freely acting asymmetric orbifold
    $\mathbb{R}^{1,5}\times T^4/\mathbb{Z}_3$, characterized by the twist vectors  $\tilde{u}=\left(-\frac{1}{3},\frac{1}{3}\right)$ and $u=\left(\frac{1}{3},\frac{1}{3}\right)$. The massless fields make up the $\mathcal{N}=4\,(1,1)$ gravity multiplet and 20 vector multiplets in $6D$.
\end{itemize}
We continue with the cusps of the modulus $U$ and we keep $T$ fixed:
\begin{itemize}
      \item $U\to i\infty$: All gravitini remain massive. The theory becomes type IIA on a non-freely acting asymmetric orbifold $\mathbb{R}^{1,5}\times T^4/\mathbb{Z}_6$, characterized by the twist vectors  $\tilde{u}=\left(\frac{5}{6},\frac{1}{6}\right)$ and $u=\left(\frac{1}{6},\frac{1}{6}\right)$. The massless fields make up the $\mathcal{N}=2$ gravity multiplet in $6D$ coupled to $9$ tensor multiplets, $8$ vector multiplets and $20$ hypermultiplets, satisfying the gravitation anomaly cancellation condition $n_H-n_V=273-29n_T$.
    \item $U\to 0$: All gravitini become massless and supersymmetry in enhanced from $\mathcal{N}=2$ to $\mathcal{N}=8$. The resulting theory is  type IIA theory on $\mathbb{R}^{1,5}\times T^4$. The massless fields make up the $\mathcal{N}=8$ gravity multiplet is $6D$.
    \item $U\to -1/2$: The two R-NS gravitini carrying momentum number $n_1 = 2\bmod 6$ and $n_1 = 4\bmod 6$ become massless and supersymmetry in enhanced from $\mathcal{N}=2$ to $\mathcal{N}=4$. At this cusp we obtain type IIA on a non-freely acting symmetric orbifold $\mathbb{R}^{1,5}\times T^4/\mathbb{Z}_2$, characterized by the twist vectors $\tilde{u}=u=\left(\frac{1}{2},\frac{1}{2}\right)$. The massless fields make up the $\mathcal{N}=4\,(1,1)$ gravity multiplet and 20 vector multiplets in $6D$. The same spectrum could also be obtained from type IIA on $\mathbb{R}^{1,5}\times \mathrm{K}3$. 
    \item $U\to -1/3$: The two R-NS gravitini carrying momentum number $n_1 = 3 \bmod 6$ become massless and supersymmetry in enhanced from $\mathcal{N}=2$ to $\mathcal{N}=4$. In this case, we get type IIA theory on a non-freely acting asymmetric orbifold
    $\mathbb{R}^{1,5}\times T^4/\mathbb{Z}_3$, characterized by the twist vectors  $\tilde{u}=\left(-\frac{1}{3},\frac{1}{3}\right)$ and $u=\left(\frac{1}{3},\frac{1}{3}\right)$. The massless fields make up the $\mathcal{N}=4\,(2,0)$ gravity multiplet and 21 tensor multiplets in $6D$, ensuring that gravitational anomalies cancel.
\end{itemize}
It is easy to see that the various cusp points of the modulus $T$ can be mapped to those of $U$ by the following transformation
\begin{equation}
   \gamma:\quad T\leftrightarrow 1/\widebar{U}\,, \qquad U\leftrightarrow 1/\widebar{T}\,. 
\end{equation}
Such a transformation is an element in $\text{O}(2,2;\mathbb{Z})$, but not in $\text{SO}(2,2;\mathbb{Z})$, so it also maps type IIB to type IIA. The construction and action on the lattice of this element is written at the end of \autoref{details on partition app}. Note that this transformation changes the chirality of the right-moving Ramond vacuum; this is important for understanding the representations of the various supergravity fields in $6D$.

\subsubsection{Double cusps as 5$D$ limits}
So far, we have discussed the single cusps on the $T-U$ plane of the moduli space. However, there are limits on the boundary of both spaces: the double cusps. The analysis of these double cusps is similar to the analysis of the single cusps. Hence, we will omit most of the details and we will simply present our results, which we collect in \autoref{doublecusps2}. Note that at all double cusps the effective supergravity theory becomes five-dimensional. Regarding our conventions for the massless spectra at each of the double cusps we refer to \cite{Gkountoumis:2023fym}.

\renewcommand{\arraystretch}{1.7}
\begin{table}[t!]
\centering
 \begin{tabular}[c]{|c|c|c|c|c|c|}
    \hline
    $T$ & $U$ & $\alpha'm^2$ & Compactification & Supersymmetry & Massless spectrum \\
    \hline
    \hline
  $ 0 $& $0$ &\multirow{9}{*}{$w_2^2T_2U_2$} &  $\left(T^4\times S^1\right)/\mathbb{Z}_6$ & $\mathcal{N}=2$ IIA & 
 $1\text{GM}+2\text{VM}+x\text{HM}$  \\
  \cline{1-2}\cline{4-6}
  $0$ & $-\frac{1}{2}$ & & $\left(T^4\times S^1\right)/\mathbb{Z}_6$, $u=\left(0,\frac{1}{3}\right)$  & $\mathcal{N}=2$ IIA & $1\text{GM}+8\text{VM}+7\text{HM}$ \\
  \cline{1-2}\cline{4-6}
  $0$ & $-\frac{1}{3}$ & & $\left(T^4\times S^1\right)/\mathbb{Z}_6$, $u=\left(0,\frac{1}{2}\right)$ & $\mathcal{N}=2$ IIA & $1\text{GM}+8\text{VM}+10\text{HM}$   \\
  \cline{1-2}\cline{4-6}
  $-2$ & $0$ & & $ \left(T^4\times S^1\right)/\mathbb{Z}_2$ & $\mathcal{N}=4\,(1,1)$ IIA & $1\text{GM}+5\text{VM}$  \\
  \cline{1-2}\cline{4-6}
  $-3$ & $0$ & & $ \left(T^4\times S^1\right)/\mathbb{Z}_3$ & $\mathcal{N}=4\,(2,0)$ IIA & $1\text{GM}+3\text{VM}$   \\
  \cline{1-2}\cline{4-6}
  $-2$ & $-\frac{1}{2}$ & & $T^4/\mathbb{Z}_2\times S^1$ & $\mathcal{N}=4\,(1,1)$ IIA & $1\text{GM}+21\text{VM}$   \\
  \cline{1-2}\cline{4-6}
  $-3$ & $-\frac{1}{3}$ & & $T^4/\mathbb{Z}_3\times S^1$ & $\mathcal{N}=4\,(2,0)$ IIA & $1\text{GM}+21\text{VM}$   \\
  \cline{1-2}\cline{4-6}
  $-2$ & $-\frac{1}{3}$ & & \multirow{2}{*}{$\left(T^4\times S^1_{\mathcal{R}_5=\sqrt{6}}\right)/\mathbb{Z}_6$} & \multirow{2}{*}{$\mathcal{N}=2$ IIA} & \multirow{2}{*}{$1\text{GM}+2\text{VM}+4\text{HM}$}  \\
  \cline{1-2}
  $-3$ & $-\frac{1}{2}$ & &  &  &    \\
  \hline
  $i\infty$ & $i\infty$ & {$\frac{n_2^2}{T_2U_2}$} & $\left(T^4\times S^1\right)/\mathbb{Z}_6$ & {$\mathcal{N}=2$ IIB} &
 $1\text{GM}+2\text{VM}+x\text{HM}$   \\
    \hline
  $i\infty$ & $0$ & \multirow{3}{*}{$\frac{n_1^2U_2}{T_2}$} & $T^5$ & $\mathcal{N}=8$ IIB & $1\text{GM}$ \\
  \cline{1-2}\cline{4-6}
  $i\infty$ & $-\frac{1}{2}$ &  &$ \left(T^4\times S^1\right)/\mathbb{Z}_2$  & $\mathcal{N}=4\,(0,2)$ IIB & $1\text{GM}+5\text{VM}$  \\
  \cline{1-2}\cline{4-6}
  $i\infty$ & $-\frac{1}{3}$ &  & $\left(T^4\times S^1\right)/\mathbb{Z}_3$  & $\mathcal{N}=4\,(1,1)$ IIB & $1\text{GM}+3\text{VM}$  \\
  \hline
  $0$ & $i\infty$ & \multirow{3}{*}{$\frac{\hat{w}_1^2T_2}{U_2}$} & $T^4/\mathbb{Z}_6\times S^1$ & $\mathcal{N}=2$ IIA & $1\text{GM}+18\text{VM}+20\text{HM}$  \\
  \cline{1-2}\cline{4-6}
  $-2$ & $i\infty$ &  & $\left(T^4\times S^1\right)/\mathbb{Z}_6$, $u=\left(0,\frac{1}{3}\right)$ & $\mathcal{N}=2$ IIA & $1\text{GM}+8\text{VM}+7\text{HM}$  \\
  \cline{1-2}\cline{4-6}
  $-3$ & $i\infty$ &  & $\left(T^4\times S^1\right)/\mathbb{Z}_6$, $u=\left(0,\frac{1}{2}\right)$  & $\mathcal{N}=2$ IIA & $1\text{GM}+8\text{VM}+10\text{HM}$  \\
   \hline
    \end{tabular}
\captionsetup{width=.9\linewidth}
\caption{Here we present our results for all double cusps in the $T$-$U$ moduli space. First, we list the double-cups. Then we specify the mass of the towers that become massless and the resulting decompactified theory at the corresponding cusp. In all cases there are five non-compact directions ($\mathbb{R}^{1,4}$) and five compact directions. Here, GM stands for gravity multiplet, VM stands for vector multiplet, and HM stands for hypermultiplet. Regarding the massless spectrum at the cusps $(T,U)\to(0,0)$ and $(i\infty, i\infty)$, if  $T_2/U_2=6$, $x=4$; if $T_2/U_2=12$, $x=3$; and for all other cases $x=2$. Finally, we specify the shift vectors for some freely acting orbifold limits, of which the shift along the circle is not inversely proportional to the orbifold rank.}
\label{doublecusps2}
\end{table}
\renewcommand{\arraystretch}{1}

There are four $T$-cusps and four $U$-cusps, and by combination there are 16 double-cusp infinite distance limits on the moduli space. The masses of the towers of states that become massless as we approach each double cusp can be derived from \eqref{mkfinal}. As an example, we consider the limit $T\to i\infty$ and $U\to -1/2$. In this case we find a KK tower with mass 
\begin{equation}\label{m=nU/T}
    m^2 = \frac{n_1^2 U_2}{\alpha' T_2} =\frac{n_1^2}{\alpha' }e^{\sqrt{2}(\phi_U-\phi_T)}\propto e^{-2|\bm{\phi}|},
\end{equation}
for $\hat{w}_1=w_2=0$, and $n_1=-2n_2$. These constraints on the lattice of momenta and windings imply that this tower appears only in the untwisted sector, that is $k=0$, and if $n_1\in 2\mathbb{Z}$. Furthermore, there are two gravitini carrying momentum $n_1\in 2\mathbb{Z}$, which become massless at this double-cusp limit. Hence, supersymmetry is enhanced from $\mathcal{N}=2$ to $\mathcal{N}=4$ in $5D$.\footnote{In our notation, $\mathcal{N}=2$ supersymmetry in 5$D$ means 8 supersymmetries.} We collect all information about the masses of the towers that become zero, the constraints on the momentum and winding numbers and the sectors in which massless towers appear at each of the 16 double cusps in \autoref{doublecusps}.

Note that at the cusps $(T,U)=(-2,-1/3),\,(-3,-1/2)$, there are two additional towers of hypermultiplets that become massless, since we are exactly at the critical line $\frac{T}{6}=U$ (cf. \eqref{Tmod6U}).  Regarding the cusps $(T,U)=(0,0)$ and $(i\infty,i\infty)$, if the  ratio $T_2/U_2$  as we approach the cusp is 12, we obtain one extra tower of massless hypermultiplets (cf.  \eqref{t12branch}), and if the ratio is 6 we get two towers of massless hypermultiplets (cf. \eqref{Tmod6U}). 

Moreover, at the cusps $(T,U)=(0,-1/2),\,(0,-1/3),\,(-2,i\infty),\,(-3,i\infty)$ the theory decompactifies to an orbifold of $\mathbb{R}^{1,4}\times S^1 \times T^4$, which acts as a rotation of order $6$ on the torus and as a shift of order $2$ or $3$ on the circle. The fact that the shift is not of order $6$ has important implications for the spectrum of the orbifold. Consider for example the case in which the rotation on the torus is of order $6$ and the shift on the circle is of order $3$. In this orbifold, the winding number along the circle direction will be shifted as $w\to w+k/3$, which implies that states in the $k=3$ twisted sector will not feel the shift. Consequently, there will be massless states coming from this twisted sector. Moreover, states with momentum number $n$ along the circle direction will pick up a phase $e^{2\pi i n /3}$. So, states with orbifold charge $e^{\pi i  /3}, e^{\pi i }$ or $e^{5\pi i /3}$ will be projected out of the spectrum, since such orbifold charge cannot be cancelled by adding momentum along the circle direction. The situation is similar if the shift along the circle is of order $2$. In this case, states in the $k=2$ and $4$ sectors will not feel the shift, and states with orbifold charge $e^{\pi i  /3}, e^{2\pi i  /3}, e^{4\pi i  /3}$ or $e^{5\pi i /3}$ will be projected out of the spectrum.

Finally, as in the example \eqref{m=nU/T}, the masses of all towers that become massless at the double cusps are proportional to both $\exp{\left(\pm\frac{1}{\sqrt{2}}\phi_T\right)}$ and $\exp{\left(\pm\frac{1}{\sqrt{2}}\phi_U\right)}$. Hence, the masses of the towers decrease exponentially with $\lambda=1$ (see \eqref{SDC}), and the SDC is verified also in the case of double cusps in the $T-U$ moduli space.

\renewcommand{\arraystretch}{1.7}
\begin{table}[t!]
\centering
 \begin{tabular}[c]{|c|c|c|c|c|c|}
    \hline
    $T$ & $U$ & $\alpha'm^2$ & Lattice constraints & $n_1(\bmod 6)$ & Massless sectors \\
    \hline
    \hline
  $ 0 $& $0$ &\multirow{9}{*}{$w_2^2T_2U_2$} & $n_1=n_2=\hat{w}_1=0$ & $0$ &$0\, (1,5)$  \\
  \cline{1-2}\cline{4-6}
  $0$ & $-\frac{1}{2}$ & & $n_1=-2n_2=0,2\hat{w}_1=w_2$ & $0$ & $0,3$ \\
  \cline{1-2}\cline{4-6}
  $0$ & $-\frac{1}{3}$ & & $n_1=-3n_2=0,3\hat{w}_1=w_2$ & $0$ & $0,2,4$   \\
  \cline{1-2}\cline{4-6}
  $-2$ & $0$ & & $2\hat{w}_1=n_2=0,n_1=-2w_2$ & $0,2,4$ & $0$  \\
  \cline{1-2}\cline{4-6}
  $-3$ & $0$ & & $3\hat{w}_1=n_2=0,n_1=-3w_2$ & $0,3$ & $0$   \\
  \cline{1-2}\cline{4-6}
  $-2$ & $-\frac{1}{2}$ & & $-n_1=2n_2=4\hat{w}_1=2w_2$ & $0,2,4$ & $0,3$   \\
  \cline{1-2}\cline{4-6}
  $-3$ & $-\frac{1}{3}$ & & $-n_1=3n_2=9\hat{w}_1=3w_2$ & $0,3$ & $0,2,4$   \\
  \cline{1-2}\cline{4-6}
  $-2$ & $-\frac{1}{3}$ & & $-n_1=3n_2=6\hat{w}_1=2w_2$ & \multirow{2}{*}{$0$} & \multirow{2}{*}{$0,1,5$}  \\
  \cline{1-2}\cline{4-4}
  $-3$ & $-\frac{1}{2}$ & & $-n_1=2n_2=6\hat{w}_1=3w_2$ &  &    \\
  \hline
  $i\infty$ & $i\infty$ & $\frac{n_2^2}{T_2U_2}$ & $n_1=\hat{w}_1=w_2=0$ & 0 & $0\,(1,5)$   \\
  \hline
  $i\infty$ & $0$ & \multirow{3}{*}{$\frac{n_1^2U_2}{T_2}$} & $\hat{w}_1=w_2=n_2=0$ & All & 0  \\
  \cline{1-2}\cline{4-6}
  $i\infty$ & $-\frac{1}{2}$ &  & $2\hat{w}_1=w_2=0,n_1=-2n_2$ & $0,2,4$ & 0  \\
  \cline{1-2}\cline{4-6}
  $i\infty$ & $-\frac{1}{3}$ &  & $3\hat{w}_1=w_2=0,n_1=-3n_2$ & $0,3$ & 0  \\
  \hline
  $0$ & $i\infty$ & \multirow{3}{*}{$\frac{\hat{w}_1^2T_2}{U_2}$} & $n_1=w_2=n_2=0$ & 0 & All  \\
  \cline{1-2}\cline{4-6}
  $-2$ & $i\infty$ &  & $n_1=-2w_2=0,n_2=2\hat{w}_1$& 0 & $0,3$  \\
  \cline{1-2}\cline{4-6}
  $-3$ & $i\infty$ &  & $n_1=-3w_2=0,n_2=3\hat{w}_1$ & 0 & $0,2,4$  \\
   \hline
    \end{tabular}
\captionsetup{width=.9\linewidth}
\caption{Here we list the masses of towers that become massless, the constraints on the lattice of momenta and windings and the sectors in which massless states appear at each double cusp. For the cusps $(T,U)\to (0,0)$ and $(i\infty,i\infty)$, there are extra towers of massless hypermultiplet arising from the $k=1$ and $5$ sectors, if $T=6U$ or $T=12U$ asymptotically.}
\label{doublecusps}
\end{table}
\renewcommand{\arraystretch}{1}

\section{Conclusion and discussion}
\label{Conclusion and discussion}

In this work we have checked that the 
distance conjecture holds for a particular
non-geometric string compactification, which is  a freely acting asymmetric $\mathbb{Z}_6$ orbifold of type IIB string theory with a classical $STU$ moduli space.
 This was a non-trivial test, as the duality group of the orbifolded theory was reduced to subgroups of the modular group due to the shift along the circle coordinate. Hence, new points of infinite distance on the real axis of the moduli space needed to be examined. We chose this particular example because of its rich structure, but we expect our conclusions to hold more generally. 

In our example, all infinite distance points corresponded to decompactification limits to either six or five dimensions. As we explicitly demonstrated by studying the orbifold partition function, at the cusps on the real axis of the moduli space a freely acting asymmetric orbifold could decompactify to a non-freely acting symmetric orbifold. Also, at some cusps, some or all gravitini became massless and supersymmetry was enhanced from $\mathcal{N}=2$ to $\mathcal{N}=4$ or $\mathcal{N}=8$.

In addition to the distance conjecture, there 
are conjectures that the volume of moduli space should be finite or that its asymptotic growth be restricted
 \cite{Ooguri:2006in}. For a recent discussion on this and the relation to dualities, see
\cite{Delgado:2024skw,Grimm:2025lip}. 
The volume of the classical moduli space 
for our model is indeed finite, 
 because the hypermultiplet moduli space is a Narain moduli space with a finite volume and the vector multiplet moduli space is the triple product of the fundamental domains of $\hat{\Gamma}^1(6)$ (or $\hat{\Gamma}_1(6)$). The index of $\hat{\Gamma}^1(6)$ is $[\mathrm{SL}(2,\mathbb{Z}):\hat{\Gamma}^1(6)]=12$ (same as $\hat{\Gamma}_1(6)$), so that the volume of the classical vector multiplet moduli space is $(12\text{Vol}(\mathbb{H}/\mathrm{SL}(2,\mathbb{Z})))^3=64\pi^3$.

Finally, we also found that a finite number of massive and charged hypermultiplets could become massless at special lines or points in the interior of the moduli space, which indicates that the classical prepotential could be modified by quantum effects\footnote{On the volume finiteness of the quantum corrected moduli space, because the geodesic distance to the singularities $\frac{T}{6}=U$ and $\frac{T}{6}=2U$ is finite, the volume also remains finite.}. The computation of such quantum corrections will be the subject of a future work.

\section*{Acknowledgements}

It is a pleasure to thank Thomas Grimm for useful discussions. The work of GN is supported by the China Scholarship Council.
The work of CH was supported by   the STFC Consolidated Grants   ST/T000791/1 and ST/X000575/1.

\appendix

\section{Details on the partition function}
\label{details on partition app}
The orbifold partition function takes the general form
\begin{equation}
    Z(\tau,\bar \tau)=\frac{1}{p}\sum_{k,l=0}^{p-1}{Z}[k,l](\tau,\bar \tau)\ ,
    \label{orbipartition1}
\end{equation}
where $\tau=\tau_1+i\tau_2$ is the complex structure modulus of the torus\footnote{The modulus $\tau$ should not be confused with the modulus $U$, which is the complex structure modulus of the background $T^2$.}, and
\begin{equation}
    {Z}[k,l]= {Z}_{\mathbb{R}^{1,3}}\,  {Z}_{T^2}[k,l]  {Z}_{T^4}[k,l]  {Z}_F[k,l]\,.
\end{equation}
Here ${Z}_{\mathbb{R}^{1,3}}$ is the contribution to the partition function from the non-compact bosons, ${Z}_{T^2}[k,l]$ and ${Z}_{T^4}[k,l]$ refer to the compact bosons on $T^2$ and $T^4$ respectively and ${Z}_F[k,l]$ is the fermionic contribution to the partition function. Recall that $p$ is the orbifold rank and $k$ labels the untwisted and twisted sectors. In addition, $l$ implements the orbifold projection in each sector. 

Also, it is useful to mention that, in general, the partition function can be factorized into left and right-moving pieces as\footnote{The bosonic zero modes along the compact directions require special treatment, as infinite sums over quantized momenta and windings may appear. The bosonic zero modes along the $T^4$ are irrelevant for our discussion but those along the $T^2$ will be discussed in detail.}
\begin{equation}
    Z[k,l] =  \widetilde{\mathcal{Z}}[\tilde{\theta}^k,\tilde{\theta}^l] \otimes \mathcal{Z}[\theta^k,\theta^l]\,.
        \label{orbipartition2}
\end{equation}
Here $\theta $ is the generator of the orbifold group; $\theta^k$ refers to twisted sectors where the torus coordinates obey boundary conditions of the form $W_{\text{R}}^i(\sigma^1,\sigma^2+2\pi)= \theta^k\, W_{\text{R}}^i(\sigma^1,\sigma^2)$  and  $\theta^l$ characterizes the orbifold action: $W_{\text{R}}^i \to \theta^l\,W_{\text{R}}^i$ ($\tilde{\theta}^k,\tilde{\theta}^l$ correspond to the left-movers).  In this appendix we will focus mostly on the behaviour of the $T^2$ partition function in the various infinite distance points. For more details on the partition function we refer to \cite{Gkountoumis:2023fym}.

Recall that the $T^2$ partition function  reads
\begin{equation}
    {Z}_{T^2}[k,l]=  \frac{1}{(\eta\widebar{\eta})^2}\sum_{\left\{n_i,w_i\right\} \in \mathbb{Z}^4}\,e^{\frac{2\pi i ln_1 }{p}} \,\widebar{q}^{\frac{1}{2}p_{\text{L}}^2(k)}\,q^{\frac{1}{2}p_{\text{R}}^2(k)}\,, \qquad i=1,2\,,
    \label{shiftedt2app}
\end{equation}
where 
\begin{equation}
    \begin{aligned}
        &p^2_{\text{L}}(k) = \frac{1}{2T_2U_2}\left|n_2-Un_1 +\widebar{T}\left(w_1+\tfrac{k}{p}\right) +\widebar{T}Uw_2\right|^2\,,\\
         &p^2_{\text{R}}(k) = \frac{1}{2T_2U_2}\left|n_2-Un_1 +{T}\left(w_1+\tfrac{k}{p}\right) +{T}Uw_2\right|^2\,.
    \end{aligned}
    \label{shifted momentaapp}
\end{equation}
We can rewrite the left and right-moving momenta in terms of the background fields of $T^2$, i.e. the metric $g_{ij}$ and antisymmetric tensor $b_{ij}$, as
\begin{equation}
\begin{aligned}
   & p_{\text{L}}^2(k) = \frac{\alpha'}{2}n_ig^{-1}_{ij}n_j+\frac{1}{2\alpha'}\hat{w}_i(g-bg^{-1}b)_{ij}\hat{w}_j+\hat{w}_i(bg^{-1})_{ij}n_j + n_i\hat{w}_i \,,\\
   & p_{\text{R}}^2(k) = \frac{\alpha'}{2}n_ig^{-1}_{ij}n_j+\frac{1}{2\alpha'}\hat{w}_i(g-bg^{-1}b)_{ij}\hat{w}_j+\hat{w}_i(bg^{-1})_{ij}n_j - n_i\hat{w}_i \,,
    \end{aligned}
    \label{momentabgtwistedapp}
\end{equation}
where, $i,j=1,2$ and summation over repeated indices is implied. Also, we have defined $\hat{w}_i \equiv w_i+k_i/p$, where $\vec{k}=(k_1,k_2)=(k,0)$. Also, let us define $\vec{l}=(l_1,l_2)=(l,0)$. Then, by performing a Poisson resummation over the momentum vector $\vec{n}=(n_1,n_2)$ we can bring \eqref{shiftedt2app} to the equivalent form
\begin{equation}
    {{Z}}_{T^2}[k,l]=\frac{\sqrt{\det g}}{\alpha'\, {\tau_2}\,(\eta\,\widebar{\eta})^2}\sum_{\left\{n_i,w_i\right\} \in \mathbb{Z}^4}e^{ \frac{-\pi}{\alpha'\tau_2}\left[n_i-\frac{l_i}{p}+\left(w_i+\frac{k_i}{p}\right)\tau\right]\left(g_{ij}-b_{ij}\right)\left[n_j-\frac{l_j}{p}+\left(w_j+\frac{k_j}{p}\right)\widebar{\tau}\right]}\,. 
    \label{shift2gb}
\end{equation}
Recall that the $d$-dimensional Poisson resummation formula is given by
\begin{equation}
     \sum_{n_i \in \mathbb{Z}^d}e^{-\pi  n_iA_{ij}n_j+\pi B_i n_i}=(\text{det}A)^{-\frac{1}{2}} \sum_{n_i \in \mathbb{Z}^d}e^{-{\pi}\left(n_i+i\frac{B_i}{2}\right)(A^{-1}){ij}\left(n_j+i\frac{B_j}{2}\right)}\,.
\end{equation}
Also,
\begin{equation}
    g_{ij} = \alpha'\frac{T_2}{U_2}\begin{pmatrix}
        1&U_1\\
        U_1 & |U|^2
    \end{pmatrix}\,,\qquad \text{and} \qquad b_{ij}= \alpha'\begin{pmatrix}
        0& T_1\\
        -T_1&0
    \end{pmatrix}\,.
    \label{metricandb}
\end{equation}
Let us now focus on the cusps of the modulus $T$, and keep the modulus $U$ fixed in the bulk. For convenience, we set $U_1=0$. Then,  $T_2=\mathcal{R}_5\mathcal{R}_4/\alpha'$ and $U_2=\mathcal{R}_4/\mathcal{R}_5$. First we study the limit $T_2\to \infty$, that is $\mathcal{R}_5\to \infty$ and $\mathcal{R}_4\to \infty$. In this limit, $g_{ij}\to \infty$ and the only term that contributes in the sum in \eqref{shift2gb} is the term with $n_1=n_2=w_1=w_2=k=l=0$. So, we find
\begin{equation}
  \lim_{T_2\to\infty}  {{Z}}_{T^2}[0,0]=  
   \frac{\mathcal{R}_4\mathcal{R}_5}{\alpha'\, {\tau_2}\,(\eta\,\widebar{\eta})^2}\,,
\end{equation}
while for $k$ and/or $l\neq 0$ the limit is exponentially suppressed. Using that the string length $\ell_s$ is given by $\ell_s=2\pi\sqrt{\alpha'}$, we can rewrite the above expression as
\begin{equation}
   \lim_{T_2\to\infty}  {{Z}}_{T^2}[0,0]=  \frac{4\pi^2\mathcal{R}_4\mathcal{R}_5}{\ell_s^2\, {\tau_2}\,(\eta\,\widebar{\eta})^2} = \frac{V}{\ell_s^2} \frac{1}{ {\tau_2}\,(\eta\,\widebar{\eta})^2}\,,
   \label{freet2}
\end{equation}
where $V$ is the volume of a very large two-torus. Now, we recognize that the expression \eqref{freet2} is the properly normalized partition function of two non-compact bosons (see e.g. \cite{Blumenhagen:2013fgp} or \cite{Kiritsis:2019npv}). Combining this result with the orbifold partition function \eqref{orbipartition1}-\eqref{orbipartition2}, we conclude that in the limit $T\to i\infty$, the resulting theory is type IIB on $\mathbb{R}^{1,5}\times T^4$.

We mention here that there is a subtlety regarding the volume of the torus that becomes very large. Since the orbifold partition function is divided by the orbifold rank $p$, the volume of the torus that decompactifies is actually $4\pi^2\mathcal{R}_4\mathcal{R}_5/p$. Moreover, it is interesting to note that the radius $\mathcal{R}_5$ on which the orbifold acts by a shift is related to the radius $R_5$ of the corresponding Scherk-Schwarz effective supergravity theory by $\mathcal{R}_5=pR_5$.

Now we focus on the limit $T_2\to 0$, i.e $\mathcal{R}_4\to 0$ and $\mathcal{R}_5\to 0$. Here, we have three different cusps, namely the cusps at $T=0,-2$ and $-3$. We start from the cusp $T_1=0$. In this case, the partition function \eqref{shift2gb} reads
\begin{equation}
    {{Z}}_{T^2}[k,l]=\frac{\mathcal{R}_4\mathcal{R}_5}{\alpha'\, {\tau_2}\,(\eta\,\widebar{\eta})^2}\sum_{\left\{n_i,w_i\right\} \in \mathbb{Z}^4}e^{ \frac{-\pi\mathcal{R}_5^2}{\alpha'\tau_2}\left|n_1-\frac{l}{p}+\left(w_1+\frac{k}{p}\right)\tau\right|^2}e^{ \frac{-\pi\mathcal{R}_4^2}{\alpha'\tau_2}|n_2+w_2\tau|^2}\,. 
    \label{factt2gb}
\end{equation}
By performing a multiple Poisson resummation over all momentum and winding numbers we can bring \eqref{factt2gb} to the equivalent form
\begin{equation}
     {{Z}}_{T^2}[k,l]=\frac{\alpha'}{\mathcal{R}_5\mathcal{R}_4}\frac{1}{\tau_2\,(\eta\,\widebar{\eta})^2}\sum_{\left\{n_i,w_i\right\} \in \mathbb{Z}^4}e^{\frac{2\pi i }{p}(n_1l+w_1k)}e^{ \frac{-\pi\alpha'}{\mathcal{R}_5^2\tau_2}|w_1+n_1\tau|^2}e^{ \frac{-\pi\alpha'}{\mathcal{R}_4^2\tau_2}|w_2+n_2\tau|^2}\,. 
     \label{2tdualities}
\end{equation}
It is easy to verify that the above result could also be obtained by performing the T-duality transformation $\mathcal{R}_5\to \alpha'/\mathcal{R}_5$, $\mathcal{R}_4\to \alpha'/\mathcal{R}_4$, $n_1\leftrightarrow w_1$, $n_2\leftrightarrow w_2$. Note that this transformation changes the shift vector $u=(1/p,0,0,0)$ to $\tilde{u}=(0,0,1/p,0)$. As we can see from \eqref{2tdualities}, in the limit $\mathcal{R}_4\to 0$ and $\mathcal{R}_5\to 0$ all terms are exponentially suppressed except for the terms with $n_1=n_2=w_1=w_2=0$ and $k,l=0,\ldots,p$. In particular, we find that for all $k,l$
\begin{equation}
     \lim_{T_2\to 0}{{Z}}_{T^2}[k,l]=\frac{\alpha'}{\mathcal{R}_5\mathcal{R}_4}\frac{1}{\tau_2\,(\eta\,\widebar{\eta})^2}\,,
\end{equation}
or, by defining the dual radii $\widetilde{\mathcal{R}}_5=\alpha'/\mathcal{R}_5$, $\widetilde{\mathcal{R}}_4=\alpha'/\mathcal{R}_4$
\begin{equation}
     \lim_{T_2\to 0}{{Z}}_{T^2}[k,l]=\frac{\widetilde{\mathcal{R}}_4\widetilde{\mathcal{R}}_5}{\alpha'\tau_2\,(\eta\,\widebar{\eta})^2}\,.
\end{equation}
So, in the limit $T\to 0$ we obtain the partition function of two non-compact bosons, for all values of $k$ and $l$. Then, from \eqref{orbipartition1}-\eqref{orbipartition2} we can see that in the limit $T\to 0$ the theory becomes type IIB on a non-freely acting asymmetric orbifold $\mathbb{R}^{1,5}\times T^4/\mathbb{Z}_6$, characterized by the twist vectors  $\tilde{u}=\left(\frac{5}{6},\frac{1}{6}\right)$ and $u=\left(\frac{1}{6},\frac{1}{6}\right)$.

We continue with the cusp $T_1=-2$, for which we use the partition function given in \eqref{shiftedt2app}-\eqref{momentabgtwistedapp}. After a bit of algebra, it is easy to verify that, in the limit $T_2\to 0$, all terms in the partition function are exponentially suppressed unless 
\begin{equation}
    n_1 = -2w_2\qquad \text{and}\qquad n_2 = 2w_1 +\frac{k}{3}\,.
    \label{t-2cusp}
\end{equation}
Here, in order to make contact with the model studied in section \ref{z12 with nv=3 and nh=0}, we have used that $p=6$. Note that \eqref{t-2cusp} has solutions only for $k=0$ and $3$. Now, by plugging \eqref{t-2cusp} back in \eqref{shiftedt2app}-\eqref{momentabgtwistedapp} we obtain
\begin{equation}
   \lim_{\substack{T_2\to 0\\ T_1=-2}}{{Z}}_{T^2}[k,l] = \lim_{T_2\to 0} \frac{1}{(\eta\widebar{\eta})^2}\sum_{\left\{w_1,w_2\right\} \in \mathbb{Z}^2}\,e^{\frac{-2\pi i lw_2 }{3}}e^{ {-\pi\alpha'\tau_2}(\hat{w}_1^2\mathcal{R}_5^2 + w_2^2\mathcal{R}_4^2)}\,.
\end{equation}
By performing a double Poisson resummation over the winding numbers $w_1$ and $w_2$ we find
\begin{equation}
    \lim_{\substack{T_2\to 0\\ T_1=-2}}{{Z}}_{T^2}[k,l] = \lim_{\substack{\mathcal{R}_5\to 0\\ \mathcal{R}_4\to 0}}\frac{\alpha'}{\mathcal{R}_5\mathcal{R}_4}\frac{1}{\tau_2\,(\eta\,\widebar{\eta})^2}\sum_{\left\{w_1,w_2\right\} \in \mathbb{Z}^2}e^{\frac{\pi i k w_1}{3}}e^{ \frac{-\pi\alpha'}{\mathcal{R}_5^2\tau_2}w_1^2} e^{ \frac{-\pi\alpha'}{\mathcal{R}_4^2\tau_2}(w_2+\frac{l}{3})^2}\,.
\end{equation}
From this expression we see that all terms are exponentially suppressed unless 
\begin{equation}
    w_1=0\qquad \text{and} \qquad w_2+\frac{l}{3}=0\,,
\end{equation}
which is satisfied only for $l=0$ and $3$. Putting everything together, we conclude that if $[k,l]=[0,0],[0,3],[3,0]$ or $[3,3]$
\begin{equation}
     \lim_{\substack{T_2\to 0\\ T_1=-2}}{{Z}}_{T^2}[k,l]=\frac{\alpha'}{\mathcal{R}_5\mathcal{R}_4}\frac{1}{\tau_2\,(\eta\,\widebar{\eta})^2}\,,
\end{equation}
while for all other values of $[k,l]$ the limit is exponentially suppressed. Now, from \eqref{orbipartition1}-\eqref{orbipartition2} we can see that we obtain type IIB on a non-freely acting symmetric orbifold $\mathbb{R}^{1,5}\times T^4/\mathbb{Z}_2$, characterized by the twist vectors  $\tilde{u}=u=\left(\frac{1}{2},\frac{1}{2}\right)$.

Finally, the analysis of the cusp $T_1=-3$ is completely analogous to the case $T_1=-2$, so we omit the details. We find that if $[k,l]=[0,0],[0,2],[0,4],[2,0],[2,2],[2,4],[4,0],[4,2]$ or $[4,4]$
\begin{equation}
     \lim_{\substack{T_2\to 0\\ T_1=-3}}{{Z}}_{T^2}[k,l]=\frac{\alpha'}{\mathcal{R}_5\mathcal{R}_4}\frac{1}{\tau_2\,(\eta\,\widebar{\eta})^2}\,,
\end{equation}
while for all other values of $[k,l]$ the limit is exponentially suppressed. In this case we obtain type IIB on a non-freely acting asymmetric orbifold
$\mathbb{R}^{1,5}\times T^4/\mathbb{Z}_3$, characterized by the twist vectors  $\tilde{u}=\left(-\frac{1}{3},\frac{1}{3}\right)$ and $u=\left(\frac{1}{3},\frac{1}{3}\right)$.

Now, we discuss the cusps of the modulus $U$ and we keep $T$ constant. Also, for convenience, we set $T_1=0$. We start from the limit $U_2\to\infty$, and without loss of generality, we set $U_1=0$. Then,  $U_2\to\infty$ implies that $\mathcal{R}_4\to \infty$ and $\mathcal{R}_5\to 0$. In order to study this limit, we start from \eqref{factt2gb} and we perform a double Poisson resummation over the momentum number $n_1$ and the winding number $w_1$. We find 
\begin{equation}
    {{Z}}_{T^2}[k,l]= \frac{\mathcal{R}_4/\mathcal{R}_5}{ {\tau_2}\,(\eta\,\widebar{\eta})^2}\sum_{\left\{n_i,w_i\right\} \in \mathbb{Z}^4}e^{\frac{2\pi i }{p}(n_1l+w_1k)}e^{ \frac{-\pi\alpha'}{\mathcal{R}_5^2\tau_2}|w_1+n_1\tau|^2}e^{ \frac{-\pi\mathcal{R}_4^2}{\alpha'\tau_2}|n_2+w_2\tau|^2}\,. 
\end{equation}
From this expression it is easy to see that for all $k,l$
\begin{equation}
     \lim_{U_2\to \infty}{{Z}}_{T^2}[k,l]=\frac{\mathcal{R}_4/\mathcal{R}_5}{ {\tau_2}\,(\eta\,\widebar{\eta})^2}\,.
\end{equation}
We continue with the limit $U_2\to 0$. Here we have three different cusps at $U_1=0$, $-\frac{1}{3}$ and $-\frac{1}{2}$. We start from the cusp 
$U_1=0$. Now, $U_2\to 0$ implies that $\mathcal{R}_4\to 0$ and $\mathcal{R}_5\to\infty$.   In order to analyse this limit we start from \eqref{factt2gb} and we perform a double Poisson resummation over the momentum number $n_2$ and the winding number $w_2$.
We obtain
\begin{equation}
    {{Z}}_{T^2}[k,l]=\frac{\mathcal{R}_5/\mathcal{R}_4}{ {\tau_2}\,(\eta\,\widebar{\eta})^2}\sum_{\left\{n_i,w_i\right\} \in \mathbb{Z}^4}e^{ \frac{-\pi\mathcal{R}_5^2}{\alpha'\tau_2}\left|n_1-\frac{l}{p}+\left(w_1+\frac{k}{p}\right)\tau\right|^2}e^{ \frac{-\pi\alpha'}{\mathcal{R}_4^2\tau_2}\left|w_2+n_2\tau\right|^2}\,. 
    \label{poison1}
\end{equation} 
From the above expression it is clear that the only term that is not exponentially suppressed is the term with $n_1=n_2=w_1=w_2=k=l=0$. Thus, we obtain
\begin{equation}
    \lim_{U_2\to 0}{{Z}}_{T^2}[0,0]=\frac{\mathcal{R}_5/\mathcal{R}_4}{ {\tau_2}\,(\eta\,\widebar{\eta})^2}\,.
    \label{dualr4}
\end{equation}
while for $k$ and/or $l\neq 0$ the limit is exponentially suppressed.

Let us now consider the limit $U_2\to 0$, with $U_1=-\tfrac{1}{2}$. In this case, $g_{ij}\to \infty$, as we can see from \eqref{metricandb}. After a bit of algebra, it is easy to see that the sum in \eqref{shift2gb} is exponentially suppressed unless the following condition is met (recall that the model of section \ref{z12 with nv=3 and nh=0} is a $\mathbb{Z}_6$ orbifold, so $p=6$):
\begin{equation}
 n_2=2n_1-\frac{l}{3}\qquad \text{and}\qquad
   w_2= 2w_1+\frac{k}{3}\,.
    \label{condition33}
\end{equation}
First of all, it is clear that for $l$ and/or $k=1,2,4$ and $5$, the condition \eqref{condition33} can never be met. Moreover, if $[k,l]=[0,0],[0,3],[3,0]$ or $[3,3]$, \eqref{condition33} fixes $n_2$ and $w_2$ in terms of $n_1$ and $w_1$, respectively. By substituting \eqref{condition33} back in  \eqref{shift2gb}, we obtain
\begin{equation}
   \lim_{\substack{U_2\to 0\\ U_1=-{1}/{2}}}{{Z}}_{T^2}[k,l]= \frac{T_2}{{\tau_2}\,(\eta\,\widebar{\eta})^2}\lim_{U_2\to 0}\sum_{n_1,w_1\in \mathbb{Z}}e^{ \frac{-\pi 4 T_2U_2}{\tau_2}\left|n_1-\frac{l}{6}+\left(w_1+\frac{k}{6}\right)\tau\right|^2}\,.
\end{equation}
By performing a double Poisson resummation on $n_1$ and $w_1$ we get
\begin{equation}
    \lim_{\substack{U_2\to 0\\ U_1=-{1}/{2}}}{{Z}}_{T^2}[k,l]= \frac{1}{{\tau_2}\,(\eta\,\widebar{\eta})^2}\lim_{U_2\to 0}\frac{1}{4U_2}\sum_{n_1,w_1\in \mathbb{Z}}e^{\frac{2\pi i }{6}(n_1l+w_1k)}e^{ \frac{-\pi}{ 4 \tau_2T_2U_2}|w_1+n_1\tau|^2}\,.
\end{equation}
So, all terms are exponentially suppressed unless  $n_1=w_1=0$, which yields
\begin{equation}
    \lim_{\substack{U_2\to 0\\ U_1=-{1}/{2}}}{{Z}}_{T^2}[k,l]=\frac{1}{4U_2} \frac{1}{{\tau_2}\,(\eta\,\widebar{\eta})^2} \,.
\end{equation}
The analysis of the limit $U_2\to 0$, with $U_1=-\tfrac{1}{3}$ and $T$ constant, proceeds in a similar way. In this case, we find that the sum in \eqref{shift2gb} is exponentially suppressed unless 
\begin{equation}
 n_2=3n_1-\frac{l}{2}\qquad \text{and}\qquad
   w_2= 3w_1+\frac{k}{2}\,.
    \label{condition22}
\end{equation}
This condition can be solved if $[k,l]=[0,0],[0,2],[0,4],[2,0],[2,2],[2,4],[4,0],[4,2]$ or $[4,4]$. For these values of $[k,l]$ we find
\begin{equation}
    \lim_{\substack{U_2\to 0\\ U_1=-{1}/{3}}}{{Z}}_{T^2}[k,l]=\frac{1}{9U_2} \frac{1}{{\tau_2}\,(\eta\,\widebar{\eta})^2} \,.
\end{equation}
We would like to mention here that the results for the cusps of the modulus $U$ can be simply obtained from those of $T$ by performing the T-duality transformation $T\to 1/\widebar{U}$. To be precise, let us denote the element of $\mathrm{O}(2,2;\mathbb{Z})$ that exchanges the moduli $T$ and $ U$ by $\gamma_e$; this element also exchanges type IIB with type IIA theory. Furthermore, we denote the coordinate reflection $Z_1\to -Z_1$, which acts on the moduli as $(T,U)\to (-\widebar{T},-\widebar{U})$, by $\gamma_r$. These two $\mathbb{Z}_2$'s act on a vector of the lattice $\Gamma^{2,2}$ as
\begin{equation}
    (T, U)\to(U,T):\quad \begin{pmatrix}
        0&0&1&0\\
        0&-1&0&0\\
        1&0&0&0\\
        0&0&0&-1
    \end{pmatrix},\qquad (T,U)\to (-\widebar{T},-\widebar{U}):\quad \begin{pmatrix}
        -1&0&0&0\\
        0&1&0&0\\
        0&0&-1&0\\
        0&0&0&1
    \end{pmatrix}.
\end{equation}
Finally, we denote the transformation $(T,U)\to (-1/T,U)$ of $\mathrm{SL}(2,\mathbb{Z})_T$ as $\gamma_i$. Then, the T-duality element $\gamma=\gamma_i\gamma_e\gamma_r\gamma_i$ act as
\begin{equation}
     T\leftrightarrow \frac{1}{\widebar{U}}, \quad U\leftrightarrow \frac{1}{\widebar{T}},\qquad \begin{pmatrix}
        w_1\\ w_2\\n_1\\n_2
    \end{pmatrix}\leftrightarrow\begin{pmatrix}
        w_1\\ n_2\\n_1\\w_2
    \end{pmatrix}\,.
\end{equation}
Note that this duality element leaves the shift vector invariant, and it exchanges the cusps of $T$ and $U$ as follows:
\begin{equation}
    \begin{aligned}
        U\to 0 \qquad&\leftrightarrow&\quad T\to i\infty\,,\\
        T\to 0 \qquad&\leftrightarrow& U\to i\infty\,,\\
        U\to -\frac{1}{2} \qquad&\leftrightarrow& T\to-2\,,\\
        U\to -\frac{1}{3} \qquad&\leftrightarrow& T\to-3\,,\\
    \end{aligned}
\end{equation}
Finally it is easy to see that $\text{det}(\gamma)=-1$. Hence, the T-duality element $\gamma$ exchanges type IIB with type IIA theory. Concluding, we can see that the cusps of the modulus $T$ can be interpreted as various decompactification limits of the type IIB theory, while the cusps of the modulus $U$ can be interpreted as various decompactification limit of the type IIA theory.

\section{Supergravity multiplets in $6D$}
\label{6dsugra}

In this appendix we discuss the various supergravity fields, and supergravity multiplets in $6D$. First of all, we list in \autoref{tablemasslessstatesapp} the weight vectors of the lightest left and right-moving states in the untwisted orbifold sector, and their representations under the massless little group $\text{SU}(2)\times \text{SU}(2)$ in $6D$.

\renewcommand{\arraystretch}{2}
\begin{table}[h!]
\centering
 \begin{tabular}{|c|c|c|}
    \hline
    Sector &  $\tilde{r}, {r}$   & $\text{SU}(2)\times \text{SU}(2)$ rep\\
    \hline
    \hline
   \multirow{3}{*}{NS}  & $(\underline{\pm 1,0},0,0)$  & $(\textbf{2},\textbf{2})$\\
    \cline{2-3}
    &$(0,0,\pm 1,0)$&  2$\,\times\,(\textbf{1},\textbf{1})$\\
    \cline{2-3}
   & $(0,0,0,\pm 1)$ & 2$\,\times\,(\textbf{1},\textbf{1})$\\
    \hline
    \hline
    \multirow{4}{*}{R}  & $(\pm\frac{1}{2},\pm\frac{1}{2},\frac{1}{2},\frac{1}{2})$ & $(\textbf{2},\textbf{1})$\\
    \cline{2-3}
    & $(\pm\frac{1}{2},\pm\frac{1}{2},-\frac{1}{2},-\frac{1}{2})$ & $(\textbf{2},\textbf{1})$\\
    \cline{2-3}
    & $(\underline{\frac{1}{2},-\frac{1}{2}},\frac{1}{2},-\frac{1}{2})$ & $(\textbf{1},\textbf{2})$\\
    \cline{2-3}
    & $(\underline{\frac{1}{2},-\frac{1}{2}},-\frac{1}{2},\frac{1}{2})$ & $(\textbf{1},\textbf{2})$\\
   \hline
    \end{tabular}
\captionsetup{width=.9\linewidth}
\caption{The weight vectors of the lightest left and right-moving states in the untwisted sector, and their representations under the massless little group $\text{SU}(2)\times \text{SU}(2)$ in $6D$. Underlying denotes permutation.}
\label{tablemasslessstatesapp}
\end{table}
\renewcommand{\arraystretch}{1}

Also, for the construction of states, we use the rule $\textbf{2}\otimes \textbf{2} = \textbf{3} \oplus \textbf{1}$, for tensoring SU$(2)$ representations.  In addition, we table the massless representations that correspond to the various supergravity fields in six dimensions in \autoref{table5Dfieldreps}.

\renewcommand{\arraystretch}{1.2}
\begin{table}[h!]
\centering
\begin{tabular}{|c|c|}
\hline
\;\,Massive field\,\; & $\text{SU}(2)\times \text{SU}(2)$ rep \\ \hline\hline
$B_{\mu\nu}^+$ / $B_{\mu\nu}^-$ & $(\textbf{3},\textbf{1})$ / $(\textbf{1},\textbf{3})$ \\
$\psi_\mu^+$ / $\psi_\mu^-$ & \;\:$(\textbf{2},\textbf{3})$ / $(\textbf{3},\textbf{2})$\:\; \\
$A_\mu$ & $(\textbf{2},\textbf{2})$ \\
$\chi^+$ / $\chi^-$ & $(\textbf{1},\textbf{2})$ / $(\textbf{2},\textbf{1})$ \\
$\phi$ & $(\textbf{1},\textbf{1})$ \\ \hline
\end{tabular}
\captionsetup{width=.83\linewidth}
\caption{{Here we show the various massless $6D$ supergravity fields and their representations under the massless little group.}}
\label{table5Dfieldreps}
\end{table}
\renewcommand{\arraystretch}{1}

Now, we can present the various supergravity multiplets.

\subsection*{$\mathcal{N}=8$ }
All massless fields fit in the gravity multiplet, in the representations
\begin{equation}
(\textbf{3},\textbf{3})\oplus4\times(\textbf{3},\textbf{2})\oplus4\times(\textbf{2},\textbf{3})\oplus 5\times(\textbf{3},\textbf{1})\oplus 5\times(\textbf{1},\textbf{3})\oplus16\times(\textbf{2},\textbf{2})\oplus20\times(\textbf{2},\textbf{1})\oplus20\times(\textbf{1},\textbf{2})\oplus25\times(\textbf{1},\textbf{1})\,.
\end{equation}

\subsection*{$\mathcal{N}=4$ $(0,2)$ }
There are two types of multiplets. The gravity multiplet
\begin{equation}
(\textbf{3},\textbf{3})\oplus4\times(\textbf{2},\textbf{3})\oplus 5\times(\textbf{1},\textbf{3})\,,
\end{equation}
and the tensor multiplet
\begin{equation}
(\textbf{3},\textbf{1})\oplus4\times(\textbf{2},\textbf{1})\oplus 5\times(\textbf{1},\textbf{1})\,.
\end{equation}

\subsection*{$\mathcal{N}=4$ $(1,1)$ }
Again, we have two types of multiplets. The gravity multiplet
\begin{equation}
(\textbf{3},\textbf{3})\oplus2\times(\textbf{3},\textbf{2})\oplus2\times(\textbf{2},\textbf{3})\oplus (\textbf{3},\textbf{1})\oplus (\textbf{1},\textbf{3})\oplus4\times(\textbf{2},\textbf{2})\oplus2\times(\textbf{2},\textbf{1})\oplus2\times(\textbf{1},\textbf{2})\oplus(\textbf{1},\textbf{1})\,,
\end{equation}
and the vector multiplet
\begin{equation} (\textbf{2},\textbf{2})\oplus2\times(\textbf{2},\textbf{1})\oplus2\times(\textbf{1},\textbf{2})\oplus 4\times(\textbf{1},\textbf{1})\,.
\end{equation}

\subsection*{$\mathcal{N}=2$ }
There exist four types of multiplets. The gravity multiplet
\begin{equation}
(\textbf{3},\textbf{3})\oplus2\times(\textbf{2},\textbf{3})\oplus (\textbf{1},\textbf{3})\,,
\end{equation}
the tensor multiplet
\begin{equation}
(\textbf{3},\textbf{1})\oplus2\times(\textbf{2},\textbf{1})\oplus (\textbf{1},\textbf{1})\,,
\end{equation}
the vector multiplet
\begin{equation} (\textbf{2},\textbf{2})\oplus2\times(\textbf{1},\textbf{2})\,.
\end{equation}
and the hypermultiplet
\begin{equation} 2\times(\textbf{2},\textbf{1})\oplus 4\times(\textbf{1},\textbf{1})\,.
\end{equation}

\bibliographystyle{JHEP}
\bibliography{bib}

\end{document}